\documentclass{jfm}
\usepackage{graphicx}
\usepackage{epstopdf, epsfig}
\usepackage{lipsum}
\usepackage{graphicx}
\usepackage{xcolor}
\usepackage{todonotes}
\usepackage{comment}
\usepackage[super]{nth}
\usepackage{amssymb}
\usepackage{amsmath}
\shorttitle{DNS of a Boundary Layer with Strong Pressure Gradients}
\shortauthor{R. Balin and K. E. Jansen}

\title{Direct Numerical Simulation of a Turbulent Boundary Layer with Strong Pressure Gradients}

\author{Riccardo Balin\aff{1}
  \corresp{\email{riccardo.balin@colorado.edu}},
 \and K. E. Jansen\aff{1}}

\affiliation{\aff{1}Ann and H. J. Smead Aerospace Engineering Sciences, University of Colorado Boulder,
Boulder, CO 80309, USA}

\begin{document}

\maketitle

%%%%%%%%%%%%%%%%%%%%%%%%%%%%%%%%%%%%%%%%%%%%%%%%%%%%%%%%%%%%%
%%%%%%%%%%%%%%%%%%%%%%%%%%%%%%%%%%%%%%%%%%%%%%%%%%%%%%%%%%%%%
%%%%%%%% Abstract %%%%%%%%%%%%%%%%%%%%%%%%%%%%%%%%%%%%%%%%%%%
\begin{abstract}
The turbulent boundary layer over a Gaussian shaped bump is computed by direct numerical simulation (DNS) of the incompressible Navier-Stokes equations. The two-dimensional bump causes a series of strong pressure gradients alternating in rapid succession. At the inflow, the momentum thickness Reynolds number is approximately $1,000$ and the boundary layer thickness is $1/8$ of the bump height.
DNS results show that the strong favorable pressure gradient (FPG) causes the boundary layer to enter a relaminarization process. The near-wall turbulence is significantly weakened and becomes intermittent, however relaminarization does not complete. The streamwise velocity profiles deviate above the standard logarithmic law and the Reynolds shear stress is reduced. The strong acceleration also suppresses the wall-shear normalized turbulent kinetic energy production rate.
At the bump peak, where the FPG switches to an adverse gradient (APG), the near-wall turbulence is suddenly enhanced through a partial retransition process. This results in a new highly energized internal layer which is more resilient to the strong APG and only produces incipient flow separation on the downstream side.
In the strong FPG and APG regions, the inner and outer layers become largely independent of each other. The near-wall region responds to the pressure gradients and determines the skin friction. The outer layer behaves similarly to a free-shear layer subject to pressure gradients and mean streamline curvature effects.
Results from a RANS simulation of the bump are also discussed and clearly show the lack of predictive capacity of the near-wall pressure gradient effects on the mean flow.

\end{abstract}

%%%%%%%%%%%%%%%%%%%%%%%%%%%%%%%%%%%%%%%%%%%%%%%%%%%%%%%%%
%%%%%%%%%%%%%%%%%%%%%%%%%%%%%%%%%%%%%%%%%%%%%%%%%%%%%%%%


%%%%%%%%%%%%%%%%%%%%%%%%%%%%%%%%%%%%%%%%%%%%%%%%%%%%%%%%%
%%%%%%%% Body of Paper %%%%%%%%%%%%%%%%%%%%%%%%%%%%%%%%%%
%%%%%%%%%%%%%%%%%%%%%%%%%%%%%%%%%%%%%%%%%%%%%%%%%%%%%%%%%
\section{Introduction}
\label{sec:Intro}
Turbulent boundary layers undergoing pressure gradients and separation have been the subject of a large number of studies %since their first detailed description by Prandtl in the early \nth{20} century 
due to their ubiquity in science and engineering. 
Indeed, a deep understanding of their complex physics is necessary for the accurate prediction and design of many engineering systems.
%lift and drag forces on vehicles, energy production, heat transfer rates between fluids and solid objects, and many more topics crucial to the efficient and safe design of engineering systems.
While still limited to low Reynolds numbers and small domains, the continuous improvement in computational resources has made investigation by direct numerical simulation (DNS) increasingly popular. 
DNS have not only been of great complement to experimental work in the study of boundary layer physics, but have also provided detailed and dense high fidelity data for the evaluation and improvement of all lower fidelity turbulence models, from large eddy simulation (LES) to Reynolds Averaged Navier-Stokes (RANS) closures \citep{Abe_DNS_2012,Coleman_DNS_2018,Matai_BumpLES_2019,BalinAviation2020}.

A significant number of DNS of boundary layers with pressure gradients and separation have focused on the flow over a flat plate. In these studies, a turbulent boundary layer is forced to undergo a region of adverse pressure gradient (APG) which causes the flow to separate, followed by a region of favorable pressure gradient (FPG) in order to reattach the flow and control the size of the separation bubble. 
%The pressure gradients are often applied by transpiration of the velocity across the top boundary of the domain. 
Pioneering work in this area was performed by \citet{Spalart_DNS_1993}, although in their case the boundary layer did not separate.
This was followed by a great number of studies, including \citet{Spalart_DNS_1997,Na_DNS_1998,Skote_DNS_2002,Manhart_DNS_2002,Abe_DNS_2012,Abe_DNS_2017,Kitsios_DNS_2016,Coleman_DNS_2018}, performing simulations of the same kind at increasingly higher values of the momentum thickness Reynolds number $Re_\theta$.
%although in their case the boundary layer did not separate,
%and the APG was preceded by a mild FPG whose purpose was to allow realistic turbulence to develop with a smaller increase in $Re_\theta$ than a zero pressure gradient (ZPG) region. 
%and was followed by Spalart and Coleman \cite{Spalart_DNS_1997} for which 
%separation did occur. 
%However, due to restricted computational resources, the Reynolds number was quite low ($Re_\theta \approx 230$) and the boundary layer turbulence was not fully developed before separation. 
%Later came the works of Na and Moin \cite{Na_DNS_1998}, Skote and Henningson \cite{Skote_DNS_2002}, Manhart and Friedrich \cite{Manhart_DNS_2002}, and Abe \textit{et al.} \cite{Abe_DNS_2012,Abe_DNS_2017}, performing simulations of the same kind at increasingly higher $Re_\theta$,
%reaching values around 1,000 at the inflow.  
%Kitsios \textit{et al.} \cite{Kitsios_DNS_2016} performed a DNS of the self-similar flat plate boundary layer under adverse pressure gradient and no separation, reaching Reynolds numbers around $Re_\theta=6,000$ in the interior of the domain. 
%More recently, Coleman \textit{et al.} \cite{Coleman_DNS_2018} computed the separated flow over a flat plate with inflow $Re_\theta$ spanning from around 900 to 1,700, and used the solution data to evaluate a number of RANS models on the same flow. 

Flat plate boundary layer flows provide valuable insight into smooth-body shallow separation caused by continued APG effects, which is a well-known deficiency of RANS closures. However, they lack two important characteristics that are often present in engineering applications, namely strong favorable pressure gradients, in particular a strong FPG upstream of the APG region, and strong streamline curvature.
The effects of these two phenomena on boundary layer turbulence were initially investigated experimentally with flows over bumps and hills. 
\citet{Tsuji_hills} reported the breakdown of the logarithmic law in the FPG region and the formation of internal layers at the locations where the pressure gradient changed sign.
%denoted by knee points or kinks in the Reynolds stresses. The origin of the internal layers was traced to the locations where the pressure gradient changed sign and they were attributed to an abrupt change of the shear stress gradient at the wall.
\citet{Baskaran_part1,Baskaran_part2} conducted a detailed investigation of streamwise pressure gradient and curvature effects for the flow over a convex hill. 
%Internal layers were also produced by this flow, however in this case they were attributed to abrupt changes in the surface curvature %at the leading and trailing edges. 
A decoupling of the inner and outer layers was observed, with the latter being isolated and behaving as a free shear layer. Curvature effects were most pronounced in the outer layer, affecting the turbulence structure and rapidly decreasing the Reynolds shear stress. Conversely, changes in pressure gradient initially appeared in the inner region.
\citet{Webster_bump} conducted experiments of the flow over a circular bump. They reported a significant deviation of the velocity in the FPG region above the logarithmic law 
%consistent with the findings of \citet{Patel_FPG} 
and suggested that the early stages of a relaminarization process were taking place.

Numerical simulations of boundary layers over hills and bumps have also been performed, however most employed LES due to the large computational cost required by these types of flows.
\citet{Wu_LESbump} simulated the experimental bump flow of \citet{Webster_bump} with wall-resolved LES at a Reynolds number almost three times lower, and confirmed 
%the presence of internal layers and 
the significant departure above the logarithmic law in the FPG region. 
This geometry was also studied by \citet{Cavar2011} and \citet{Matai_BumpLES_2019} with LES producing similar results. \citet{Matai_BumpLES_2019} expanded the study by considering a family of bumps of increasing height. For all cases, a large departure of the velocity above the logarithmic law was observed in the FPG as well as a plateau or a rapid oscillation in the skin friction at the start of the APG.
They showed that the non-dimensional favorable pressure gradient
\begin{equation}
\Delta_p= \frac{\nu}{\rho u_\tau^3} \frac{\partial p}{\partial s}
\label{eq:DeltaP}
\end{equation}
exceeded the value of -0.018 identified by \citet{Patel_FPG} as the start of a relaminarization process. Note that in \eqref{eq:DeltaP}, $\nu$ is the kinematic viscosity, $\rho$ is the fluid density, $p$ is the static pressure at the wall, $u_\tau$ is the friction velocity, and $s$ is the streamwise direction.
\citet{Uzun_NASAHumpLES_2018} performed a wall-resolved LES study of the NASA wall-mounted hump and compared the simulation results to the experiments of \citet{Greenblatt_NASAhump}. Both numerical and experimental data showed the presence of a plateau in the skin friction coefficient profile on the upstream side of the bump where the FPG was strong. It was noted that in this region the boundary layer was undergoing a relaminarization process according to the acceleration parameter based on the edge velocity $U_e$
\begin{equation}
K= \frac{\nu}{U_e^2}\frac{\partial U_e}{\partial s}
\label{eq:K}
\end{equation}
surpassing the limit of $3\times10^{-6}$ \citep{Narasimha_Relaminarization_1979}, but full relaminarization was not achieved.
%Other notable simulations are the ones by Spalart \textit{et al.} \cite{Spalart_BJWMLES} and by Uzun and Malik \cite{Uzun_BJLES_2018} of the Bachalo-Johnson transonic bump with shock-induced separation \cite{B-J_bump}. In this case the favorable pressure gradient was relatively weaker and did not cause deviations of the velocity profile from the logarithmic law.

Although often not the focus of the aforementioned studies on bump and hill flows, all reported a region of strong FPG with large deviations from standard turbulent boundary layer behavior. Some of them even mentioned signs of relaminarization (also known as reverse transition).
\citet{Patel_FPG} was among the first to report on these effects, noting a deviation of the streamwise velocity profile above the logarithmic law for large enough negative values of the pressure gradient parameter $\Delta_p$. They supposed the breakdown of the law was due to the process of reversion to laminar flow and proposed a tentative critical value of $\Delta_p=-0.018$. However, they noted that $\Delta_p$ did not describe the near-wall flow completely.
In a later study, \citet{Patel_ReverseTrans_1968} expanded the initial work and proposed the use of a different quantity to measure the effects of a favorable pressure gradient on the inner layer. They argued that the non-dimensional shear stress gradient 
\begin{equation}
\Delta_\tau=\frac{\nu \alpha}{\rho u_\tau^3}
\label{eq:DeltaTau}
\end{equation}
is more universal and identified a critical value of $-0.009$ for the beginning of the departure above the logarithmic law and reverse transition. Note that in \eqref{eq:DeltaTau}, $\alpha$ is the gradient of the total shear stress in the wall-normal direction $n$ across the viscous sublayer such that $\tau=\tau_w+\alpha n$. They further argued that the acceleration parameter $K$ is not an adequate quantity since it is not able to describe the near-wall flow directly, and based on the analysis on $\Delta_\tau$, they proposed a critical value on the pressure gradient of $\Delta_p=-0.024$.
\citet{Bradshaw1969} later corrected the critical value for $\Delta_\tau$ to $-0.013$ and noted that it marks the beginning of the log law overshoot and not necessarily the start of a relaminarization process, however it is an adequate indicator that the process is imminent.
%Badri Narayanan and Ramjee \cite{BadriNarayanan1969} suggested a value around $\Delta_p=-0.02$ and pointed out that the log-law breakdown coincided with the minimum shape factor.
\citet{Narasimha1973} and the subsequent review article \citet{Narasimha_Relaminarization_1979} also noted that the departure from the logarithmic law occurs upstream of the reverse transitional process. They proposed critical values of $\Delta_p=-0.025$ and $K=3 \times 10^{-6}$ as further indication of relaminarization, as well as a local minimum in the shape factor. These critical values on the pressure gradient were supported by a DNS of sink flows \citep{Spa86}. %with $K$ ranging between $1.5 \times 10^{-6}$ and $3.0 \times 10^{-6}$ \citep{Spa86}. 
%Moreover, \cite{Narasimha_Relaminarization_1979} described the skin friction response to a strong FPG to have an initial rise due to the flow accelerating, followed by a sharp fall and rise after the flow returns to fully turbulent, which is in agreement with experiments and simulations over bumps and hills \cite{Baskaran_part1,Webster_bump,Wu_LESbump,Matai_BumpLES_2019}.
\citet{Warnack1998} analyzed the Reynolds stresses of a boundary layer during reverse transition. They observed a decrease of the shear and streamwise stresses when normalized by the local friction velocity that persisted until the local minimum in the Reynolds number $Re_\theta$, and then a sharp rise due to retransition to turbulence. Similar behavior was shown for the shear and streamwise stress production when normalized by inner units.

Despite the APG region and separation receiving most of the attention in the literature, particularly in the context of improving RANS closures, it will become apparent throughout this paper that a complete understanding and accurate modeling of the FPG region, which may include a relaminarization process, are critical for the prediction of certain aerodynamic flows such as the one studied here. 
%For instance, relaminarization can be sought out and designed for in order to perform flow control. The reduction in wall shear stress that comes from the departure from fully-turbulent to quasi-laminar flow can be used to diminish viscous drag, whereas the retransition process after relaminarization that energizes the boundary layer can be used to lessen or even prevent flow separation. 
However, current RANS and near-wall models for LES perform adequately in mild FPG, but often significantly overpredict the skin friction and shear stress in strong FPG \citep{Matai_BumpLES_2019,BalinAviation2020,Uzun_NASAHumpLES_2018}.
Additionally, a relatively limited number of studies, especially numerical ones, are available discussing the details of turbulence undergoing strong acceleration, relaminarization, and retransition (the process describing the return to turbulence after relaminarization) over complex geometries. 
%As outlined above, a thorough literature review was not able to identify a consensus on the parameter and its critical values that should be used to fully characterize the onset, duration, and end of reverse transition.
The need for additional investigations into strong FPG flows, therefore, certainly exists and DNS are especially well-suited for this purpose since so much of the FPG effects are located very near the wall, offering an advantage over experimental data.

The work presented in this paper offers a novel case of a boundary layer undergoing strong pressure gradients and streamline curvature effects of alternating sign with the purpose of increasing the understanding of these types of flows and provide detailed data for the improvement of lower fidelity turbulence models.
A DNS of the turbulent boundary layer flowing over a Gaussian bump is performed. The flow is accelerated on the upstream side by a strong FPG, and then is quickly decelerated on the downstream side by an APG leading to incipient separation. 
At the relatively low Reynolds number chosen, the FPG acting over an extended length ($\approx 20$ boundary layer thicknesses $\delta$) causes the onset of relaminarization and a significant weakening of the near-wall turbulence. 
%Streamwise velocity profiles in this region deviate significantly above the logarithmic law and resemble a laminar flow solution in the inner layer. 
The boundary layer, however, does not relaminarize completely and stays intermittently turbulent. At the peak of the bump, the weakened near-wall turbulence experiences a sudden enhancement in intensity due to partial retransition, which in turn leads to an atypical skin friction response and a more resilient boundary layer.
The DNS was designed to focus on the part of the flow leading up to incipient separation rather than downstream of it, with additional emphasis on the FPG region, and the discussion of results reflects this choice. 
%Analysis of the turbulent flow is presented through wall and integral quantities, velocity and Reynolds stress profiles at a number of locations, contours of instantaneous vorticity at different heights across the boundary layer, as well as production rates of the turbulent kinetic energy and Reynolds shear stress.

It must be mentioned that the present DNS is part of a larger joint study with other research groups and a recent publication on this flow already exists. However, key differences are present between this paper and \citet{Uzun_SpeedBumpDNS_2020}, making them distinct reports of the flow.
In particular, this paper focuses on the strong FPG effects and discusses the flow physics involved in light of the body of literature on relaminarization and strongly accelerated flows, while \citet{Uzun_SpeedBumpDNS_2020} takes a more holistic approach with a greater emphasis on the incipient separation and subsequent recovery.
%For instance, the method used to inject unsteady flow into the domain differs for the two DNS, and so does the location of the inflow plane upstream of the bump. More importantly, perhaps, is the focus of the two papers. Here, the focus is on the strong FPG effects and to discuss the flow physics involved in light of the body of literature on relaminarization and strongly accelerated flows. By contrast, in \citet{Uzun_SpeedBumpDNS_2020} the authors take a more holistic approach examining results from a larger portion of the flow, with a greater emphasis on the incipient separation and subsequent recovery.

This paper is organized as follows. ~\textsection~\ref{sec:NumApproach} describes in detail the flow problem chosen and the numerical approach taken to obtain a solution with DNS. \textsection~\ref{sec:Results} presents and discusses the results obtained from the DNS. Finally, \textsection~\ref{sec:Conc} offers some concluding remarks.

\section{Numerical Setup}
\label{sec:NumApproach}

%%%%%%%%%%%%%%%%%%%%%%%%%%%%%%%%%%%%%%%%%%%%%%%%%%%%%%%
%%%%%%%%%%%%%%%%%%%%%%%%%%%%%%%%%%%%%%%%%%%%%%%%%%%%%%%
%%%%%%%%%%%%%%%%%%%%%%%%%%%%%%%%%%%%%%%%%%%%%%%%%%%%%%%
\subsection{Problem Definition}
\label{sec:NumApproachDef}
\begin{figure}
\centering
  \includegraphics[width = 0.97\textwidth]{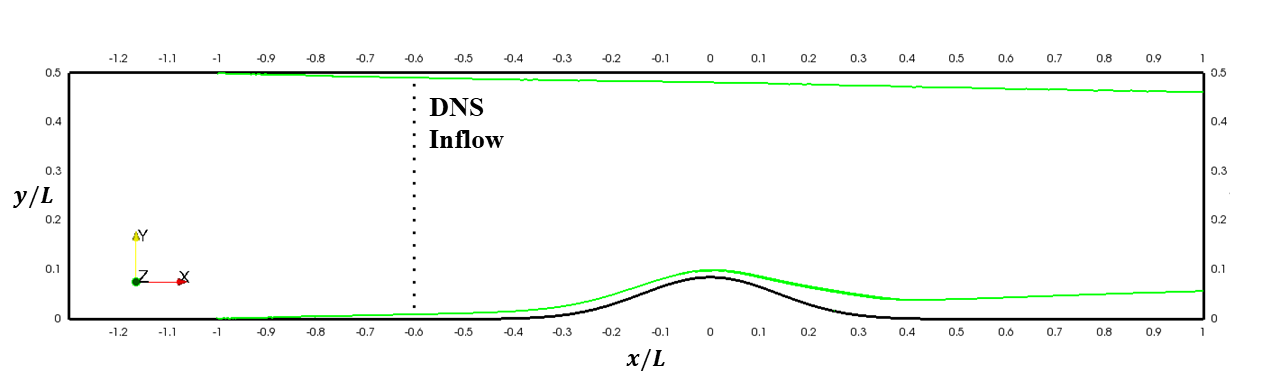}
  \caption{The black curves outline the domain of the bump flow, while the green curves show the boundary layer thickness on both no-slip walls predicted by preliminary RANS. The dotted vertical line marks the location of the inflow to the DNS.}
\label{fig:GaussBump}
\end{figure}

The flow computed in this study is the turbulent boundary layer over the prismatic extrusion of a two-dimensional (2D) Gaussian shaped bump. The surface is defined by \eqref{eq:GaussBump}, which depends on the height parameter $h$ and the length parameter $x_0$, and is shown by the black curve on the lower surface of the domain in figure~\ref{fig:GaussBump}.
\begin{equation}
y(x)=h\exp{\Big(-\big(x/x_0 \big)^2 \Big)}
\label{eq:GaussBump}
\end{equation}
Note that the $x$ coordinate is aligned with the freestream flow far upstream of the bump, the $y$ coordinate is vertical and normal to the freestream, and the $z$ coordinate is aligned with the spanwise direction. Moreover, \eqref{eq:GaussBump} defines the entire lower surface of the domain, meaning that there is no flat-plate region on either side of the bump and the curvature is everywhere continuous. 
It is important to mention that the geometry selected for this DNS is exactly the centerline of a three-dimensional (3D) bump developed at The Boeing Company \citep{Slotnik_SpeedBump_2019} and studied experimentally at the University of Washington \citep{Williams_ExpSpeedBump}. %This other bump flow, however, exhibits some three-dimensional effects that are not present with the prismatic extrusion, particularly on the downstream side of the bump and inside the separation bubble. 
Comparisons between the experimental results and the DNS, however, are deferred to future work.
%Comparisons between the experimental results and the DNS are not discussed in this paper, but this exercise is left as future work as the data from the ongoing wind tunnel campaign becomes more abundant.

%Nevertheless, some preliminary comparisons between DNS and experiments are discussed in Section~\ref{sec:Results} given the available data at the time of writing.
%For this reason, and due to the lack of detailed flow data at the time of writing, comparisons between experiments and the DNS will not be discussed in this paper (\textcolor{blue}{$C_p$ at the centerline maybe?}).
% Actually would be good to put some experimental data points for the Cp plot. Maybe even compare the velocity profile upstream of the bump to see if comparisons are even appropriate and evaluate the validity of the results relative to a true BL. If match is not so good, can say that DNS came first so there was no opportunity for corrections based on experiments. Ask Williams for data, saying what I will be using it for, and see if they have any more I can use.

To maintain similarities between the two geometries, the 2D Gaussian bump
%flow was purposefully designed to share many characteristics with the 3D flow. The 
height and length parameters were matched at $h/L=0.085$ and $x_0/L=0.195$, where $L=3$ ft is the length of the square cross-section of the wind tunnel used for the 3D bump experiments. These values were selected in order to obtain the desired pressure gradients and separation using preliminary RANS simulations of the 3D bump \citep{Slotnik_SpeedBump_2019}. The flow studied has a Reynolds number of $Re_L=1.0 \times 10^6$, corresponding to $Re_h=85,000$ when measured against the bump height. The freestream velocity is $U_\infty=16.40$ m/s, which at standard sea level conditions results in a small enough Mach number ($M_\infty=0.045$) to treat this flow as incompressible.
At the location of the inlet to the DNS, shown by the dotted vertical line in figure~\ref{fig:GaussBump}, the momentum thickness Reynolds number is $Re_\theta=1,050$ and the boundary layer thickness is approximately $1/8$ of the bump height. Due to the boundary layer being thinner than the bump height, the flow is more characteristic of a hill rather than other bumps present in the literature \citep{Webster_bump,Greenblatt_NASAhump}, however the term ``bump'' is retained.

Other boundary conditions defining the entire RANS flow domain in figure~\ref{fig:GaussBump} are as follows. 
Two solid no-slip walls are present; the bump on the lower surface defined entirely by \eqref{eq:GaussBump} and a flat top surface located at $y/L=0.5$.
The streamwise location of the origins of the top and bottom boundary layers is at the leading edges of the no-slip walls at $x/L=-1$ \citep{Slotnik_SpeedBump_2019,Williams_ExpSpeedBump}.
These are accurately simulated by extending the top and bottom surfaces upstream of the start of the no-slip walls and symmetry boundary conditions are prescribed there. This allows a uniform velocity in the $x$ direction equal to $U_\infty$ to be correctly imposed at the inflow ($x/L\approx-1.3$).

%%%%%%%%%%%%%%%%%%%%%%%%%%%%%%%%%%%%%%%%%%%%%%%%%%%%%%%
%%%%%%%%%%%%%%%%%%%%%%%%%%%%%%%%%%%%%%%%%%%%%%%%%%%%%%%
%%%%%%%%%%%%%%%%%%%%%%%%%%%%%%%%%%%%%%%%%%%%%%%%%%%%%%%
\subsection{Solution Approach}
\label{sec:NumApproachSol}
While figure~\ref{fig:GaussBump} describes the entire RANS flow domain, only a fraction of it was included in the DNS. The inflow was moved downstream to $x/L=-0.6$ as shown by the dotted line in figure~\ref{fig:GaussBump}. 
%This was done to avoid the very expensive computation of the boundary layer origin and laminar-to-turbulent transition. 
Moreover, the top wall was slanted according to a profile fitted to the displacement thickness of the boundary layer computed with a preliminary RANS simulation of the entire flow domain (see figure~\ref{fig:GaussBump}). This was done to reproduce the constriction effects of the boundary layer growing on the top wall. The preliminary RANS was carried out using the Spalart-Allmaras (SA) one-equation model \citep{spalart1994one} augmented with the rotation and streamline curvature (SARC) correction \citep{Spalart_SARC,Shur_SARC} on the full domain. Additionally, a low Reynolds number modification proposed in \citet{Coleman_DNS_2018} 
%and refined in \cite{Spalart_LowReCorr_2020} 
was applied to correct the underprediction of the skin friction coefficient by the SA model at low $Re_\theta$. 
The spanwise period of the DNS domain was set to $4.7\delta_{995_{in}}$, where $\delta_{995_{in}}$ is the inflow 99.5\% boundary layer thickness. 
Due to the significant growth of the boundary layer on the downstream side of the bump as shown in figure~\ref{fig:GaussBump}, this period may introduce some confinement effects in this part of the solution \citep{Coleman_DNS_2018,Abe_DNS_2017}. However, the region of interest for this study is the portion of the flow leading up to incipient separation, with particular emphasis on the FPG and the bump peak, where the boundary layer growth relative to the inflow is only moderate. Separation and the re-development of the boundary layer are not discussed in this paper.
%Previous DNS \cite{Coleman_DNS_2018,Abe_DNS_2017} indicated a domain period between three and four times the local boundary layer thickness is needed to avoid constraint of the large scale structures that form during separation and persist after reattachment.

The influence of the spanwise period and possible confinement effects on the flow statistics were investigated using two wall-modeled LES (WMLES) of the same flow described here (same domain and boundary conditions). Only the width of the domain was changed, with one WMLES having the same period as the DNS and the other increased the width by a factor of two. The details of the WMLES with the narrower domain (same as the DNS) are discussed in \citet{BalinAviation2020}, wherein it was showed that, while the near-wall region is poorly predicted by the wall model, the outer layer Reynolds stresses and boundary layer thickness in the region of interest of the flow are captured accurately.
Since the spanwise period affects the largest scales of the domain, which are well captured in both simulations, this exercise is representative of the confinement effects experienced by the DNS. No significant differences were observed in the flow and turbulent quantities of interest (skin friction, pressure gradient, velocity, and Reynolds stresses) obtained with the two WMLES. Consequently, the solution in the region of interest of this DNS is considered to be free of any confinement effects from the spanwise period chosen.
%However, inspection of two-point spanwise correlations and instantaneous flow structures demonstrated that confinement effects are small in the portion of the boundary layer leading up to separation for which boundary layer growth is only moderate. 
% Quantify confinement effects. 1) what has been deemed sufficient in the literature for spanwise period? Based on this, what part of the domain can we use with confidence? 2) it was a compromise for us because it was a collaboration to people had to agree based on their availability to resources
% Abe 2017 - width is 3-4 times delta_max
% CRS 2018 - width is 4 times delta_max
% Na and Moin 1998 - 50x inflow displacement thickness
% Skote - 80x inflow displacement thickness
% My DNS goes to a min of 2.5x delta and drops below 3 in the region of interest from x/L=-0.4 to -0.12. Quite a big chunk of the region. This probably means I need to look at come 2 point correlations around x/L=-0.25. In terms of displacement thickness, it is between 13 and 50. In terms of momentum thickness, it is between 20 and 70.

The boundary conditions enforced in the DNS were as follows. 
The bump surface was treated as a no-slip wall. 
The top surface was modeled as an inviscid wall offset by the RANS predicted displacement thickness described above with zero transpiration (zero velocity component normal to the surface) and zero traction. 
At the outflow, weak enforcement of zero pressure was applied along with zero traction. Effects from this boundary condition on the interior domain are contained within a streamwise distance of one local boundary layer thickness and thus did not affect the upstream solution. 
At the inflow, the synthetic turbulence generator (STG) of \citet{Shur_STG} was selected to introduce unsteady flow into the domain,
which has been shown to produce realistic turbulence a short distance downstream of the inlet for both wall-modeled LES \citep{Shur_STG} and DNS \citep{Spalart_BJWMLES}. 
%According to this approach, velocity fluctuations are first computed from a superposition of spatiotemporal Fourier modes with random amplitudes and phases. These are then scaled by prescribed profiles of the time-averaged Reynolds stresses in order to obtain the desired second order moments and thus introduce synthetic scales with both anisotropy and inhomogeneity. The fluctuations are finally added to a known mean velocity profile to obtain a time and spatially varying boundary condition. 
Note that this approach differs significantly from the one in \citet{Uzun_SpeedBumpDNS_2020} wherein a recycling method was used with the inlet further upstream at $x/L=-0.8$.
The mean Reynolds stress and velocity profiles required by the STG method were extracted from an additional RANS simulation at the inflow location of $x/L=-0.6$. This simulation used the same turbulence model (SARC-lowRe) and setup described above with only the following exception to the domain described by figure~\ref{fig:GaussBump}.
The top boundary was no longer flat and modeled with a no-slip condition, but instead it was slanted and modeled as an inviscid wall to match the DNS domain. This was
done intentionally to extract profiles along the entire height of the inflow plane consistent with the DNS domain and boundary conditions. Moreover, it was used to verify that the constriction effects were appropriately captured by the slanted upper surface, and that was indeed the case.

\begin{figure}
\centering
  \includegraphics[width = 0.97\textwidth]{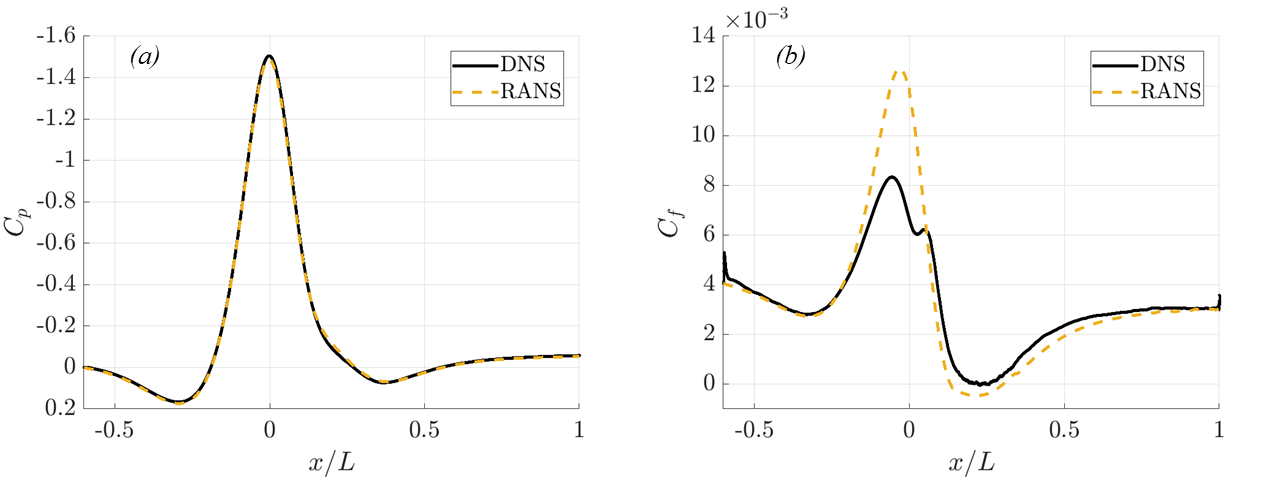}
  \caption{Pressure coefficient (a) and skin friction coefficient (b) on the surface of the  Gaussian bump.}
\label{fig:CpCf}
\end{figure}

The computational grid used for the DNS of the bump was structured with a total of 554 million points. 
It was designed with spacing $\Delta s^+=15$, $\Delta z^+_\text{max} < 10$, $\Delta n^+_1=0.1$, and $\Delta n^+_\text{max}<10$, where $(s,n,z)$ is the bump aligned coordinate system ($s$ and $n$ are tangent and normal to the bump surface, respectively). Moreover, the $()^+$ superscript signifies scaling by wall units using the friction velocity $u_\tau$ and the viscous length scale $l_\nu=\nu/u_\tau$. Note that since the DNS solution was not available at the time of the grid generation, the $u_\tau$ profile from the preliminary RANS was used (see figure~\ref{fig:CpCf} for differences in wall friction between RANS and DNS). The growth factor in the wall-normal direction was limited to $5$\%. 
The streamwise spacing $\Delta s^+=15$ was achieved everywhere by using the local value of the RANS friction velocity, however the growth or decay across adjacent elements was limited to 1\% resulting in a smooth streamwise variation of $\Delta s$. Close to the separation region predicted by RANS, where $u_\tau$ becomes ill-defined, an effective friction velocity was used instead to maintain adequate spacing in physical units. 
The wall spacing $\Delta n^+_1=0.1$ was set in a similar manner. 
On the other hand, constrained by the structured nature of the grid, $\Delta z^+$ was fixed using the maximum friction velocity over the bump surface, which occurs near the bump peak. Since the peak skin friction is overpredicted by RANS (see figure~\ref{fig:CpCf}), the maximum spanwise spacing in the DNS is actually $\Delta z^+_\text{max}=8$.
Similarly, the wall-normal spacing $\Delta n^+_\text{max}<10$ was driven by the maximum boundary layer thickness also computed by the preliminary RANS, which as shown in figure~\ref{fig:GaussBump} is found at the outflow of the domain. 
As a result, spanwise and wall-normal spacing finer than $8$ and $10$ plus units, respectively, are present over the regions discussed in this paper.
For instance, $\Delta z^+ \approx 6$ and  $\Delta n^+_\text{max} \approx 4$ in the vicinity of the inflow and up to the start of the strong FPG, and $\Delta n^+_\text{max} < 7$ everywhere upstream of incipient separation. 
%\textcolor{blue}{Should I include a few plots here to show the spacing in the region of interest in plus units? Uzun does and their grid is a finer so it might bring up questions we don't want.}

The DNS was initialized from an instantaneous solution of a wall-modeled LES simulation computed on the same domain with the same boundary conditions \citep{BalinAviation2020}. Integration of the flow was carried out for one full domain flow-through time before accumulation of statistics to ensure that the remnants of the initial condition were flushed out of the domain.
Past the transient phase, time- and span-averaged statistics were accumulated for a total time $T$ long enough to satisfy $T > 100t_\text{eddy}$ everywhere in the region of interest of this bump flow, 
%(which corresponds to 1.2 domain flow-through times), 
where the eddy-turnover time is defined based on the local edge velocity $U_e$ and boundary layer thickness $\delta$ as $t_\text{eddy}=U_e/\delta$.
Stationarity of the time- and span-averaged statistics was verified by comparing flow and turbulence quantities obtained from different time windows within the time interval $T$.
The non-dimensional time step size was $\Delta t^+=0.11$ based on the maximum average friction velocity of the DNS, which ensured a maximum Courant–Friedrichs–Lewy (CFL) number below one at each time step.

%%%%%%%%%%%%%%%%%%%%%%%%%%%%%%%%%%%%%%%%%%%%%%%%%%%%%%%
%%%%%%%%%%%%%%%%%%%%%%%%%%%%%%%%%%%%%%%%%%%%%%%%%%%%%%%
%%%%%%%%%%%%%%%%%%%%%%%%%%%%%%%%%%%%%%%%%%%%%%%%%%%%%%%
%\subsection{Flow Solver}
%\label{sec:NumApproachCode}
%The PHASTA flow-solver was used for all simulations presented in this work. PHASTA uses a stabilized, semi-discrete finite element method to solve either the compressible or incompressible Navier-Stokes equations plus any additional set of 
%scalar equations for turbulence modeling. 
%While high-order hierarchical bases of polynomials are available, tri-linear hexahedral elements were selected for this simulation resulting in second order accuracy.
%tabilization is performed with the streamline upwind/Petrov-Galerkin (SUPG)
%method \citep{Whiting1999}.
%Time integration is performed with the fully implicit, second-order accurate generalized-$\alpha$ method \citep{Jansen_GenAlpha_2000}. 
%First-order accurate time integration is also possible with the implicit Backward Euler method. 
%PHASTA was designed to work on unstructured meshes. 
%These high-fidelity, massively parallel computations could be performed in a reasonable time frame thanks to the extreme scalability of PHASTA, 
%which has been demonstrated in \citet{Sahni_PHASTAScaling_09}.
All simulations presented in this work were performed with a stabilized finite element method \citep{Whiting1999} using tri-linear hexahedral elements and second order accurate, fully implicit time integration \citep{Jansen_GenAlpha_2000}.
The accuracy of DNS with this method has been shown for a channel flow in \citet{Trofimova_DNS_2009}, in which tri-linear hexahedral elements were also used. 
Stabilization and time integration parameters (which affect numerical dissipation) chosen for this DNS follow the work of \citet{Trofimova_DNS_2009} and were verified with a flat plate DNS to reproduce skin friction, velocity, and Reynolds stresses from a number of previous studies \citep{Coles1962,Schlatter2009,Smits1983,Jimenez_ZPGDNS}.
The STG method selected for the bump was used at the inlet of the same flat plate study and was confirmed to reproduce validation data within 5 inflow boundary layer thicknesses \citep{WrightSTGDNS_arxiv}.

\section{Results and Discussion}
\label{sec:Results}

%The discussion of results of the DNS of the Gaussian bump are divided into five sections. 
%First, wall and integral quantities are presented over the entire region of interest of the domain in order to provide a global view of this boundary layer flow. Recall that the region of interest and focus of this paper extends from the inflow at $x/L=-0.60$ to incipient separation around $x/L=0.20$. 
%Second, further analysis of the flow is divided into three segments: 1) the initial mild APG present upstream of the bump ($-0.60 \le x/L < -0.29$), 2) the strong favorable pressure gradient located on the upstream face of the bump ($-0.29 \le x/L < 0.00$), and 3) the bump peak and the strong APG on the downstream side ($0.00 \le x/L \le 0.10$). Subdivision into these three segments was done for ease of presentation, however it will become clear that upstream effects are very important in this flow making the segments highly interconnected.
%Finally, production rates of the turbulent kinetic energy and Reynolds shear stress are discussed at a few key streamwise locations of this boundary layer.

%%%%%%%%%%%%%%%%%%%%%%%%%%%%%%%%%%%%%%%%%%%%%%%%%%%%%%%
%%%%%%%%%%%%%%%%%%%%%%%%%%%%%%%%%%%%%%%%%%%%%%%%%%%%%%%
%%%%%%%%%%%%%%%%%%%%%%%%%%%%%%%%%%%%%%%%%%%%%%%%%%%%%%%
% Inflow and incoming BL
%\subsection{Inflow Boundary Layer}
%\label{sec:ResInflow}
%\input{inflow.tex}

% Cf and Cp
\subsection{Wall and Integral Quantities}
\label{sec:ResCpCf}
% Discussion of time averaged wall quantities, mainly Cp and Cf

Time- and span-averaged pressure and skin friction coefficient profiles on the surface of the Gaussian bump obtained from the DNS are presented in figure~\ref{fig:CpCf}. 
The coefficients are defined in \eqref{eq:CpCf}, where $p_\text{ref}$ is the reference wall pressure at location $x/L=-0.60$ and $\tau_w$ is the wall shear stress rotated to the curvilinear coordinates $(s,n,z)$.
\begin{equation}
C_p=\frac{p_w-p_\text{ref}}{\frac{1}{2}\rho_\infty U_\infty^2}
\qquad
\qquad
C_f=\frac{\tau_w}{\frac{1}{2}\rho_\infty U_\infty^2}
\label{eq:CpCf}
\end{equation}
In this figure, the solution from the preliminary RANS simulation used to obtain the inflow profiles is also shown for comparison. 
%Recall, the turbulence model used was SA with streamline curvature and low Reynolds number corrections, and the domain for this simulation is outlined in figure~\ref{fig:GaussBump} with the exception of the slanted upper surface. 
%\textcolor{blue}{Planning to add some experimental data points for Cp.}

Very good agreement is obtained for the pressure coefficient, indicating the successful choice of the DNS sub-domain and boundary conditions.
The $C_p$ profile also outlines the series of pressure gradients experienced by the boundary layer. At the inflow to the DNS ($x/L=-0.60$), a small but adverse pressure gradient is present, indicating that the presence of the bump is already felt at that location. As the flow approaches the bump, the APG strengthens but remains mild, until $x/L=-0.29$ where it switches to a strong FPG accelerating the flow over the upstream side of the bump. At the bump peak, a rapid change from strong favorable to strong adverse occurs, with the latter persisting until about $x/L=0.40$. Finally, a mild FPG helps the boundary layer recover after incipient separation.

Significant differences are observed for the skin friction coefficient predictions in figure~\ref{fig:CpCf}. 
Fair agreement is only obtained in the initial mild APG and soon after the start of the strong FPG the curves deviate with RANS largely overpredicting $C_f$ over the bump. The DNS solution exhibits a much smaller peak and a local minimum-maximum immediately downstream of the switch from FPG to APG. This feature of the wall shear stress is also present in the DNS of \citet{Uzun_SpeedBumpDNS_2020} and is similar to the ones documented for other bump flows with strong FPG \citep{Matai_BumpLES_2019,Narasimha_Relaminarization_1979,Warnack1998}. 
Of interest is also the streamwise position of the skin friction maximum, which in both cases is located upstream of the bump peak, but occurs further upstream in the DNS. Moreover, the DNS exhibits a much steeper decrease in wall shear between the location of $C_f$ maximum and $x/L=0.00$, with a reduction of 25\% compared to 9\% for RANS.
The size of the separation region is overestimated by RANS and DNS predicts only incipient separation with the mean $C_f$ approaching zero ($C_f < 1 \times 10^{-4}$ over $0.19 < x/L < 0.27$) but only becoming negative over a short distance. Instantaneously, however, the flow at the wall does reverse direction and small confined separation bubbles are present. 
%These results are in general agreement with the DNS of \cite{Uzun_SpeedBumpDNS_2020} of the same Gaussian bump.

\begin{figure}
\centering
  \includegraphics[width = 0.97\textwidth]{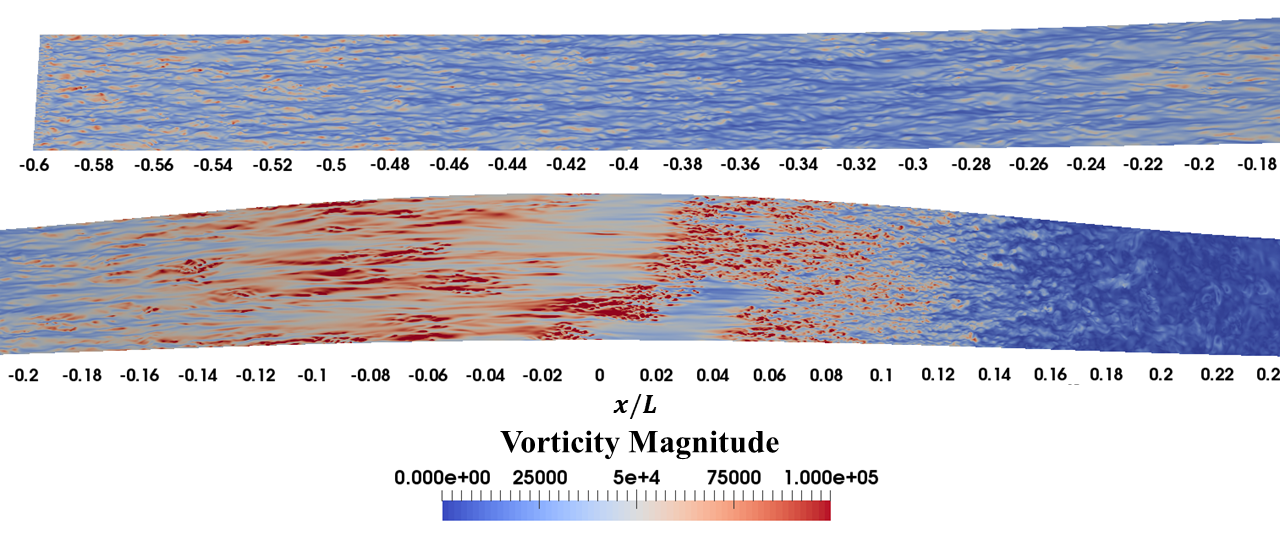}
  \caption{Contours of instantaneous vorticity magnitude on the surface of the Gaussian bump.}
\label{fig:wallVort}
\end{figure}

Contours of the instantaneous vorticity magnitude on the surface of the bump in figure~\ref{fig:wallVort} elucidate some of the features of the DNS skin friction coefficient profile.
On the upstream side of the bump, the footprint of typical near-wall structures is seen as streamwise elongated streaks of high and low vorticity. As the boundary layer progresses through the strong FPG, the acceleration of the flow causes the wall vorticity to rise and the streaks to grow significantly both in width and length.
Towards the end of the FPG, around $x/L=-0.03$ the wall vorticity drops and the streaks become weaker. Quiet regions of low vorticity also form. This behavior explains the steep drop in $C_f$ between the maximum value of the curve and the bump peak.
Starting slightly upstream of $x/L=0.00$ and continuing into the APG until approximately $x/L=0.05$, spots of large vorticity are found intermittent with the quiet regions, which visually resemble those characteristic of laminar-to-turbulent transition. Note that $x/L=0.05$ is the location of the small local maximum in the skin friction profile. These spots then appear to culminate in a region of intense turbulent activity ($0.06 \le x/L \le 0.10$) with many small-scale and fairly isotropic structures of large vorticity. Continuing further downstream, the intensity of the small-scale structures decreases as the skin friction also drops. Note that, due to the strong APG which quickly brings the flow to incipient separation, the canonical streaks do not reappear on the downstream side of the bump and by $x/L=0.15$ the turbulent structure at the wall changes dramatically once again.

Due to the strong pressure gradients and geometry of the bump, the freestream (irrotational) flow is distorted and highly non-uniform ($\partial \overline{u}_s / \partial n \ne 0$). As a result, the classical definitions of the boundary layer thickness $\delta_{995}$ and integral quantities such as the displacement and momentum thicknesses, $\delta^*$ and $\theta$ respectively, are no longer applicable. To resolve this issue, the definitions based on the generalized velocity (or ``pseudo-velocity''), $\tilde{U}$, of \citet{Spalart_DNS_1993} were used instead. These are repeated in \eqref{eq:genVel}-\eqref{eq:theta} from \citet{Coleman_DNS_2018} for convenience, where $\overline{\omega}_z$ is the mean spanwise vorticity. Note that the bump-aligned curvilinear coordinate system $(s,n,z)$ is used for the definitions instead of the freestream-aligned Cartesian system $(x,y,z)$.
\begin{equation}
\tilde{U}(s,n) \equiv - \int_{0}^{n} \overline{\omega}_z(s,n') dn'
\label{eq:genVel}
\end{equation}
\begin{equation}
\tilde{\delta}^*(s) \equiv \frac{-1}{\tilde{U}_e(s)} \int_{0}^{\infty} n\overline{\omega}_z(s,n) dn
\label{eq:dStar}
\end{equation}
\begin{equation}
\tilde{\theta}(s) \equiv \frac{-2}{(\tilde{U}_e(s))^2} \int_{0}^{\infty} n\tilde{U}(s,n)\overline{\omega}_z(s,n) dn - \tilde{\delta}^*(s)
\label{eq:theta}
\end{equation}
The edge velocity in the context of $\tilde{U}$ is defined as 
\begin{equation}
\tilde{U}_e(s) \equiv \tilde{U}(s,n \rightarrow \infty),
\label{eq:edgeVel}
\end{equation}
from which the $99.5$\% boundary layer thickness, $\tilde{\delta}_{995}$, is computed as the height above the wall where $\tilde{U}=0.995\tilde{U}_e$. This approach has been used successfully in other studies of pressure gradient flows \citep{Coleman_DNS_2018,Uzun_SpeedBumpDNS_2020}, but still requires some verification that the integrals to $n \rightarrow \infty$ are evaluated correctly. Of course, the integrals must be carried out over a finite height and it is thus important to ensure that the desired quantities are sufficiently converged under changes to the truncation. 
Integration to $1.2$ and $1.6$ times $\tilde{\delta}_{995}$ showed negligible difference in the integral quantities, and thus the latter was used.
%It was determined that integration to $\tilde{\delta}_{995}$ was not sufficient, however extending the integrals past $1.2\tilde{\delta}_{995}$ resulted in negligible differences. Consequently, the integrals were evaluated until approximately $1.6\tilde{\delta}_{995}$ for this paper, which coincides with the distance from the wall where the grid is allowed to coarsen in the wall-normal direction in order to spare resources in the freestream flow.

\begin{figure}
\centering
  \includegraphics[width = 0.97\textwidth]{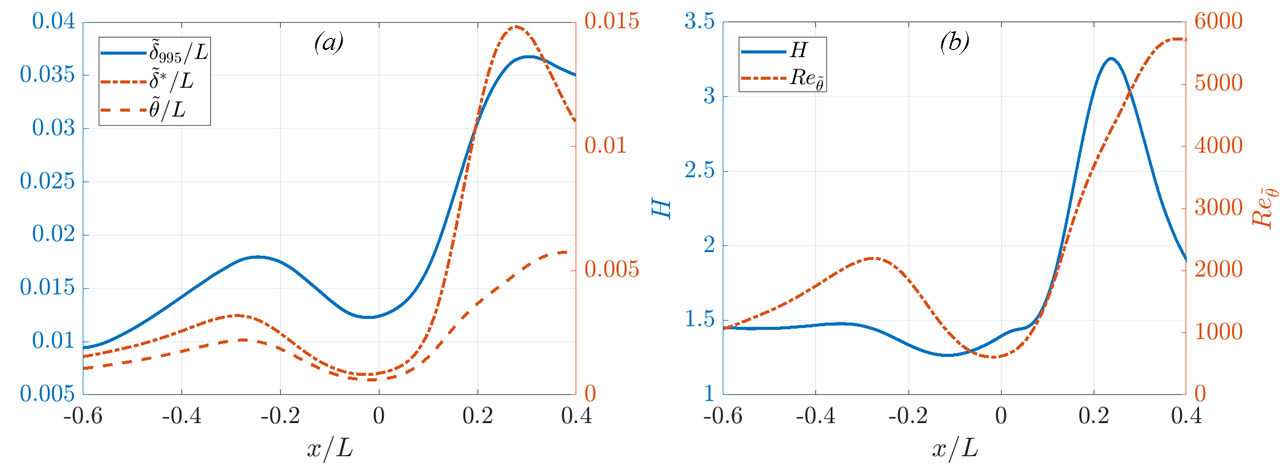}
  \caption{Variation of different boundary layer thickness measures (a) and shape factor $H$ and momentum thickness Reynolds number $Re_{\tilde{\theta}}$ (b) over the bump.}
\label{fig:BLdelta}
\end{figure}

Figure~\ref{fig:BLdelta} shows the variation of the boundary layer integral quantities defined above over the bump surface. The streamwise extent of the domain is limited to $-0.6 \le x/L \le 0.4$ in order to focus on the region of interest. 
In panel (a), the three measures of the boundary layer thickness are presented. Note that non-dimensionalization by $L$ was used. 
The $99.5$\% boundary layer thickness generally follows the trends set by the pressure gradients, with a slight delay in its response. In the initial mild APG, $\tilde{\delta}_{995}$ grows almost by a factor of two until $x/L=-0.25$ which is downstream of the start of the FPG at $x/L=-0.29$. It then shrinks as it progresses through the rest of the FPG reaching a local minimum just upstream of the bump peak. The strong APG on the downstream side of the bump causes it to grow rapidly once again creating another local maximum at $x/L=0.30$ downstream of incipient separation where $\tilde{\delta}_{995}$ is almost a factor of four larger than the inflow value.
The displacement thickness follows similar trends, however it responds more quickly to changes in pressure gradient. The first local maximum is located at $x/L=-0.29$, immediately downstream of the first change in sign of the pressure gradient. It is then reduced significantly in the strong FPG, dropping below the inflow value, and reaching a local minimum close to the bump peak at $x/L=-0.025$. In the strong APG, $\tilde{\delta}^*$ grows very rapidly to a local maximum which is almost 10 times larger than the inflow value and is also located downstream of incipient separation.
The third measure of thickness, $\tilde{\theta}$, behaves similarly to the other two. The first local maximum is located slightly downstream of the start of the FPG at $x/L=-0.27$, suggesting a slower response to the pressure gradient relative to $\tilde{\delta}^*$. The local minimum is immediately upstream of the bump peak with a value below that of the inflow. In the strong APG, $\tilde{\theta}$ also increases, although not as significantly as the displacement thickness.

Figure~\ref{fig:BLdelta}(b) shows the shape factor $H=\tilde{\delta}^*/\tilde{\theta}$. After the inflow, $H$ settles at a value of $1.44$ and then grows slightly to a local maximum of $1.48$ as the boundary layer advects through the mild APG. These are both typical values for turbulent boundary layers. The local maximum in this case is located at $x/L=-0.35$, which is well upstream of the switch from APG to FPG. The shape factor is then seen to drop through the start and middle sections of the FPG until a local minimum with value $1.27$. Note that the streamwise location of this extremum is at $x/L=-0.11$, which is significantly upstream of the bump peak. 
%The implications of this result and its relations to the reverse transitional process are discussed in Section~\ref{sec:ResFPG}. 
During the remainder of the FPG, $H$ grows and forms a plateau just downstream of the bump peak where the oscillation in $C_f$ is also observed. Finally it grows significantly in the strong APG on the downstream side of the bump until another local maximum located inside the region of incipient separation. After incipient separation, $H$ drops rapidly as the new boundary layer develops.

The momentum thickness Reynolds number, also shown in figure~\ref{fig:BLdelta}(b), follows the shape of the momentum thickness, however it is interesting to note the values of the Reynolds number in the different regions of the flow. Initially $Re_{\tilde{\theta}}$ grows from the inflow value of $1,050$ to $2,200$ at the same location of the local maximum in $\tilde{\theta}$. Just upstream of the bump peak, the minimum value of $600$ is achieved, which is well above the lowest value identified for turbulence to exist. Downstream of incipient separation, the maximum value is around $5,600$.

Streamline curvature effects are present in this flow. In the concave curvature region upstream of the bump, the non-dimensional value $\hat{\kappa}=\kappa \tilde{\delta}_{995}$ reaches a maximum value of $0.033$ at $x/L=-0.25$. Note that $\kappa$ in this context is the surface curvature and not the streamline curvature since the latter varies with distance from the wall. At the peak of the bump, where the curvature is convex, a maximum of $0.055$ is observed. Finally, in the concave region downstream of the bump $\hat{\kappa}=0.065$ at $x/L=0.26$ due to the boundary layer thickness growing significantly. As discussed in \citet{Baskaran_part2,Narasimha_Relaminarization_1979,Schwarz1996,SoMellor_Convex_73,SoMellor_Concave_75}, streamline curvature effects are mainly observed in the outer region of the boundary layer while the inner region is not directly affected unless the non-dimensional curvature $\kappa^+=\kappa \nu / u_\tau$ is large. This parameter reaches an absolute maximum of $7.9 \times 10^{-5}$ at the peak of the bump, indicating that direct curvature effects on the near-wall turbulence can be considered negligible throughout the flow leaving pressure gradient effects to be dominant.

% mild APG
\subsection{Mild Adverse Pressure Gradient}
\label{sec:ResmAPG}
% Fairly quick discussion of the mild APG effects since they are not so mild and have an influence on the BL at the start of the FPG

The initial disturbance experienced by the boundary layer is due to a weak to mild adverse pressure gradient. 
This force gradually increases from the inflow to a peak at $x/L=-0.38$ with strength $K=-0.5 \times 10^{-6}$ (see figure~\ref{fig:pGrad}), and then terminates at $x/L=-0.29$ where the pressure gradient changes sign. Concave streamline curvature effects are also present, gradually increasing from negligible values at the inflow to $\hat{\kappa}=0.028$ at the end of the APG.
Both APG and concave curvature effects are destabilizing to the boundary layer turbulence, increasing the intensity of the turbulent fluctuations and the Reynolds shear stress.

\begin{figure}
\centering
  \includegraphics[width = 0.97\textwidth]{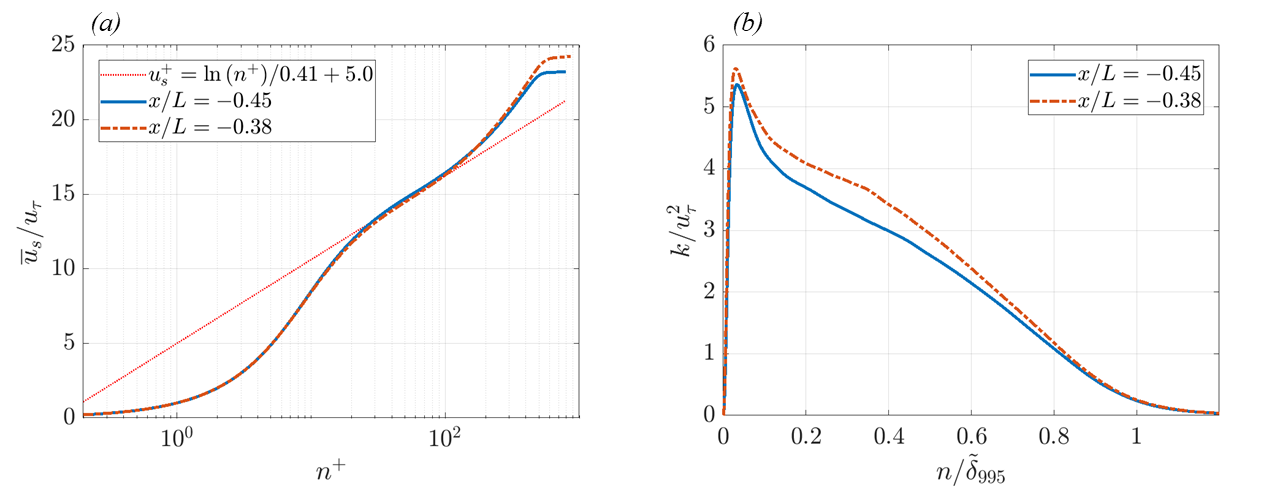}
  \includegraphics[width = 0.97\textwidth]{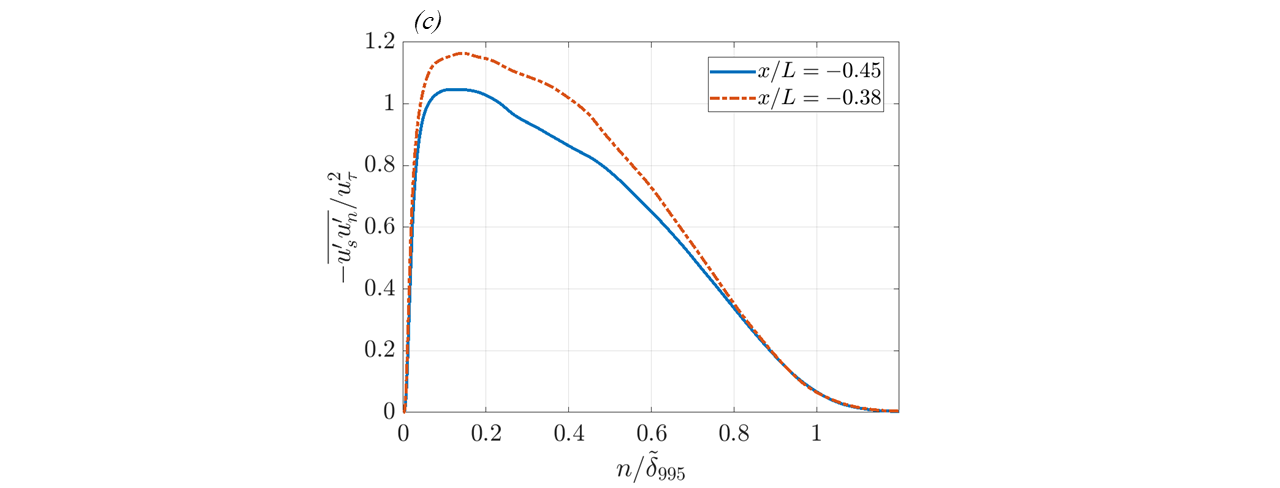}
  \caption{Profiles of the streamwise velocity (a), turbulent kinetic energy (b), and Reynolds shear stress (c) in the initial mild APG region.}
\label{fig:mildAPG}
\end{figure}

Changes to the streamwise velocity, turbulent kinetic energy (TKE), and Reynolds shear stress due to the mild APG are shown in figure~\ref{fig:mildAPG}. The TKE is defined as 
$k= \frac{1}{2} \left(  \overline{u^{\prime^{2}}_s} + \overline{u^{\prime^{2}}_n} + \overline{u^{\prime^{2}}_z} \right)$. 
Further, note that at $x/L=-0.38$, $\hat{\kappa}=0.01$. 
The velocity exhibits an increase in the wake region and a slight shortening of the logarithmic region which also appears to have a slightly lower value of the intercept. Since concave curvature causes a reduction in the wake \citep{SoMellor_Concave_75}, these affects are attributed to the APG and the increase in $Re_{\tilde{\theta}}$.
The TKE profile non-dimensionalized by the local friction velocity shows an increase mainly between $0.1 \le n/\tilde{\delta}_{995} \le 0.8$, whereas the peak value increases only marginally.
The non-dimensional Reynolds shear stress is also increased by the combined APG and curvature effects, but in this case the effects are evident everywhere below $n/\tilde{\delta}_{995}<0.8$. 

Figure~\ref{fig:mildAPG} shows that while the initial pressure gradient is only mild, its effects combined with the concave curvature cause significant changes to the boundary layer structure. Consequently, the boundary layer entering the strong FPG further downstream is not the canonical ZPG flat plate boundary layer. This feature distinguishes this bump flow from many of the experimental studies performed on strong FPG effects and relaminarization which employed a monotonic acceleration and enables the analysis of a more realistic series of pressure gradients.
%% Tsiji found that APG-FPG makes deviation from log law quicker, but deviation from the log law agrees well for me.

% FPG
\subsection{Strong Favorable Pressure Gradient}
\label{sec:ResFPG}
% This section discusses the results in the FPG region, looking at velocity, stresses but also wall and integral quantities.

This section focuses on the favorable pressure gradient region of the flow and describes the effects of this force on the mean flow and turbulence.
However, before presenting the results of this study, it is insightful to review the full spectrum of FPG affects on turbulent boundary layers which have been studied in the literature. Not all the cases discussed are present in this flow, but such a discussion grounds the observations that follow in the available body of work.
For mild streamwise gradients, these effects are well-known. Initially, they are concentrated in the inner region, causing a thickening of the viscous sublayer and an increase in the streamwise Reynolds stress as well as the wall shear stress due to the flow acceleration \citep{Baskaran_part2,Finnicum1988,Shin2015}. Subsequently, both the velocity wake factor and the velocity deficit are reduced as a result of the decay of the Reynolds shear stress in the outer region, which in turn cause a decrease in the displacement and momentum thicknesses as well as the shape factor $H=\delta^*/\theta$. Overall, an FPG has a stabilizing effect on the boundary layer turbulence. Furthermore, the streamwise velocity may still show good agreement with the standard (or a slightly displaced) logarithmic law \citep{Smits1983,Spa86}.

Under strong favorable pressure gradients, however, the effects are amplified and the flow physics are more complex.
As outlined in \citet{Sreenivasan82}, the continued action of a strong negative streamwise gradient causes the boundary layer to undergo three successive stages that significantly change its fundamental characteristics. 
Under monotonic acceleration from a ZPG canonical boundary layer, the initial response follows the trends outlined above for weak and mild FPG. The extent of this region, however, can be very short depending on the rate of change of the pressure gradient. 
As the acceleration grows stronger, significant departures occur from the standard laws but the boundary layer remains fully turbulent. Following the terminology of \citet{Sreenivasan82}, a boundary layer in this stage is said to be laminarescent. It can be considered a precursor to relaminarization, however it does not guarantee the onset of the reverse transitional process. Only the continued action of a strong acceleration will start the process. Some of the key signs of laminarescence are a deviation of the streamwise velocity from the standard logarithmic law (first below and then above \citep{Patel_ReverseTrans_1968}), significant growth of the viscous sublayer, a reduction in the $99.5$\% boundary layer thickness, and a reduction in the rate of wall-layer bursting. 

Under further sustained acceleration, the boundary layer enters a relaminarization (or reverse transitional) process. This is the second stage, and lasts until the process is complete resulting in a quasi-laminar boundary layer. It is important to highlight that this is a gradual process, thus the onset does not imply that a quasi-laminar state is achieved. During relaminarization, the previously fully turbulent boundary layer develops a viscous dominated inner layer stabilized by the acceleration, while the turbulence in the outer layer is distorted. Moreover, the skin friction is observed to drop while the flow still accelerates, the shape factor increases, and the relative (non-dimensionalized by local $u_\tau$) Reynolds stresses and TKE production drop significantly. 

The final stage occurs when the FPG is relaxed. The acceleration can no longer stabilize the near-wall flow, and soon after the onset of instability a retransition process originating near the wall returns the boundary layer towards a fully turbulent state. This process is rapid and shares some similarities to laminar-to-turbulent transition, such as the formation and growth of turbulent spots \citep{Blackwelder72}. Other notable features of this process are a sudden rise in skin friction, a significant enhancement of the turbulent intensities, and a local maximum of the shape factor.

Since relaminarization is a process that brings about significant changes to the boundary layer, it is of great interest to identify the onset and completion points. The latter can be easily defined as the location where the effects of the Reynolds stresses on the mean flow dynamics are negligible. Fluctuations are still present as a remnant from the upstream flow, but do not contribute to the development of the mean velocity. The near-wall bursting process has also ceased, therefore eliminating production of turbulent kinetic energy. The onset point, however, is more complicated to define and identify. Some previous studies \citep{Sreenivasan82,Warnack1998}, in fact, argued that it is an ill-defined exercise to determine a parameter and its critical value that uniquely define the onset because the process is not catastrophic and sudden but rather it is gradual. Nevertheless, all the non-dimensional parameters and methods proposed in the literature can be used as guidelines to suggest whether or not relaminarization has ensued. 
%It is important to mention that a degree of uncertainty is present for all of them, especially when the upstream history of the boundary layer is different relative to the published cases and when multiple disturbances are acting on the boundary layer turbulence simultaneously.

Mean streamline curvature effects are also discussed in this section.
The curvature of the Gaussian bump changes sign from concave to convex at $x/L=-0.14$, which is approximately in the middle of the FPG region. Moreover, the strongest concave effects as measured by $\hat{\kappa}$ are also located within this region at $x/L=-0.25$. As a result, both destabilizing and stabilizing effects are present from this force and are interacting with the FPG.
Concave curvature is known to increase the Reynolds shear stress and wall-normal turbulent transport of momentum, as well as reducing the wake of the streamwise velocity \citep{SoMellor_Concave_75}. 
By contrast, convex curvature can cause the Reynolds shear stress to vanish or change sign and the turbulent kinetic energy production to ``turn off'' further from the wall \citep{SoMellor_Convex_73}. Additionally, it reduces the extent of the logarithmic law and increases the wake. 
Decoupling of the inner and outer layers has been observed for boundary layers under streamline curvature of both kind, with the effects mainly focused in the outer layer and negligible impact in the near wall region, although the skin friction is subject to change (convex curvature reduces $C_f$ and vice versa) \citep{Baskaran_part2,Schwarz1996,SoMellor_Convex_73,SoMellor_Concave_75}. 
The combined effects of convex curvature and those of a weak and mild favorable pressure gradients were studied by \citet{Schwarz1996}. They noted that the pressure gradient dominates on the streamwise velocity by reducing the wake and significantly increasing the skin friction. Moreover, in the outer layer, the two stabilizing forces compound to further suppress the Reynolds shear stress, while the FPG alone reduces the relative shear throughout the inner layer.

\begin{figure}
\centering
  \includegraphics[width = 0.97\textwidth]{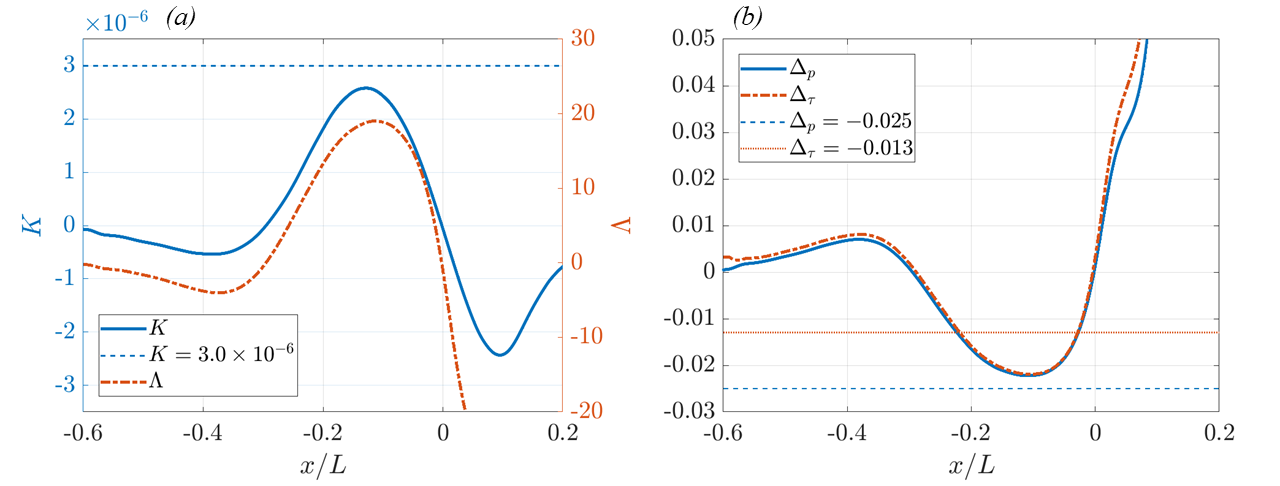}
  \caption{Pressure gradient parameters $K$ and $\Lambda$ (a) and pressure gradient $\Delta_p$ and shear stress gradient $\Delta_\tau$ non-dimensionalized by inner units (b) over the bump.}
\label{fig:pGrad}
\end{figure}

% K
Various measures of the pressure gradient are presented in figure~\ref{fig:pGrad} for the region of interest of the Gaussian bump. These, plus other instantaneous and integral quantities, will be used in a holistic approach to assess this FPG boundary layer relative to the understanding of relaminarization outlined above.
In panel (a), the acceleration parameter $K$ defined in \eqref{eq:K} is shown. Since this quantity is proportional to the gradient of the edge velocity $\tilde{U}_e$, a negative value corresponds to a deceleration and thus an APG, while a positive value corresponds to an acceleration and thus an FPG.
At the inlet, $K$ is a small negative value, indicating the presence of a very weak APG at this location. This is followed by a mild APG ($K>-1\times 10^{-6}$) until $x/L=-0.29$ and then a strong FPG over the upstream side of the bump which approaches, but does not cross, the limit of $K=3.0\times 10^{-6}$ marking the start of a reverse transitional process as outlined in previous studies \citep{Narasimha_Relaminarization_1979,Sreenivasan82,Spa86}. The maximum value is in fact $2.58\times 10^{-6}$ at $x/L=-0.13$. At the bump peak, the pressure gradient becomes adverse again, reaching a maximum strength in terms of the acceleration parameter of $-2.44\times 10^{-6}$. According to this quantity, therefore, onset of relaminarization is not achieved, even considering the range $2.75\times 10^{-6} \le K \le 3.00\times 10^{-6}$ identified with DNS of a series of sink flows \citep{Spa86}.
However, one must point out that $K$ has been criticized in previous studies \citep{Patel_ReverseTrans_1968,Sreenivasan82} for its lack of knowledge of the near-wall physics. 
% and thus lack of accuracy for flows with different types of wall boundary conditions. 

% Lambda
The non-dimensional pressure gradient
\begin{equation}
\Lambda=-\frac{\partial p_w}{\partial s} \frac{\tilde{\delta}_{995}}{\tau_w}
\label{eq:Lambda}
\end{equation}
introduced by \citet{Narasimha1973} is also shown in figure~\ref{fig:pGrad}(a). 
%It is the ratio of the pressure force acting across the boundary layer over the wall shear force. 
This parameter has been used to measure where pressure gradient forces dominate over the shear forces, thus useful in identifying the onset and completion of relaminarization \citep{Narasimha1973,Narasimha_Relaminarization_1979,Sreenivasan82}. 
Values above $50$ have been proposed as a sign of the completion of the process and the achievement of a quasi-laminar state. 
Furthermore, based on the data presented in these studies, values of $\Lambda$ between $10$ and $25$ can be correlated to the departure from fully turbulent flow and thus the onset of relaminarization.
The maximum value achieved in the FPG of the bump flow is around $19$, indicating that this boundary layer is far from reaching the completion of relaminarization, but the process is likely underway.

% Delta_p and Delta_tau
Other measures of the pressure gradient and its effects on the near-wall flow are shown in figure~\ref{fig:pGrad}(b). These are $\Delta_p$ and $\Delta_\tau$ defined in \eqref{eq:DeltaP} and \eqref{eq:DeltaTau}, respectively. Not surprisingly, the two quantities are similar, with a small offset in the mild APG region and then good agreement in the FPG. Note that due to the non-dimensionalization by the friction velocity $u_\tau$, these two quantities become ill-defined in the vicinity of incipient separation at $x/L=0.19$.
In the figure are also shown two horizontal lines, one at $\Delta_\tau=-0.013$ to mark the start of the deviation above the logarithmic law (discussed in detail later in this section) \citep{Patel_ReverseTrans_1968,Bradshaw1969} and one at $\Delta_p=-0.025$ to mark the onset of relaminarization \citep{Narasimha_Relaminarization_1979,Spa86}. 
The non-dimensional shear stress gradient profile for the bump crosses the critical value of $-0.013$ at $x/L=-0.22$ relatively soon after the start of the FPG, but neither curve crosses the limit for relaminarization. The minimum value of $\Delta_p$ is $-0.022$ at $x/L=-0.11$.
Similarly to the profile for the acceleration parameter $K$, these parameters suggest that the favorable pressure gradient is indeed strong, enough to expect the breakdown of the logarithmic law, but not enough for a relaminarization process to start. 
It is worth mentioning, however, that \citet{BadriNarayanan1969} reported $\Delta_p=-0.020$ as a critical value, which would imply that the bump boundary layer does enter a relaminarization process at $x/L=-0.17$. 
The initial study by \citet{Patel_FPG} even suggested $\Delta_p=-0.018$, although this was only a tentative value and was later updated to $-0.024$ \citep{Patel_ReverseTrans_1968}.
This variation of the critical value for $\Delta_p$ in the literature highlights the uncertainty associated with marking the onset of relaminarization for general flows with parameters based on the pressure gradient.

% H
Integral quantities have also been investigated for strongly accelerated flows in connection to the onset of relaminarization \citep{Narasimha1973,Warnack1998,Blackwelder72,Cal08}. 
Combining the known observations that the shape factor $H$ decreases for turbulent boundary layers under FPG and that laminar boundary layers exhibit much larger values of $H$, the appearance of a local minimum in this quantity followed by a sharp rise while the flow is still being accelerated has been used to determine when the boundary layer departs from a fully turbulent state to approach a quasi-laminar one. Minimum values have been observed around $1.2$ \citep{Blackwelder72}.
Similarly, \citet{Cal08} argued that the ratio $H^*=\tilde{\delta}^*/\tilde{\delta}_{995}$ initially decreases in an FPG but then suddenly grows at the onset of relaminarization, therefore making a local minimum of $H^*$ a marker of reverse transition.
For the Gaussian bump, the local minimum in $H$ is co-located with the peak of the FPG as measured by $\Delta_p$ and does not grow rapidly downstream of this point. Moreover, the minimum value of $H$ is larger than what was reported in previous studies. The local minimum of $H^*$ is found just upstream of the bump peak where $\tilde{\delta}^*$ is also a minimum and the FPG is very weak. Consequently, these integral parameters do not appear to be insightful in describing the onset of relaminarization for this flow. The uncertainties associated with using $H$ for this purpose were also discussed in \citet{Sreenivasan82}.
%Figure~\ref{fig:shapeFact} shows a close up view of these two quantities over the upstream side of the bump. A local minimum is observed for $H$ with a value of 1.27, and is co-located with the peak FPG as measured by $\Delta_p$. After the minimum, the shape factor rises to values similar to the ones at the inflow. 
%\begin{figure}
%\centering
%  \includegraphics[width = 0.45\textwidth]{figures/DNS_H-dStarOdelta.png}
%  \caption{Shape factor $H$ and boundary layer thickness ratio $\tilde{\delta}^*/\tilde{\delta}_{995}$ over the upstream side of the bump.}
%\label{fig:shapeFact}
%\end{figure}

% Cf
Another quantity not directly dependent on the pressure gradient that may be useful is the skin friction coefficient. 
A sudden decrease in the wall shear stress producing a local maximum while the flow is still undergoing acceleration has been identified as a sign of the the relaminarization process taking effect in the near-wall region \citep{Warnack1998,Narasimha1973}. The reasoning being that as the near-wall region progresses from a fully turbulent to a quasi-laminar state, the skin friction drops consistently with the known properties of laminar boundary layers. 
As shown in figure~\ref{fig:CpCf}, the local maximum is located upstream of the bump peak at $x/L=-0.055$ where the flow is still experiencing a fairly strong acceleration ($K=1.7 \times 10^{-6}$). This result is in contrast to what is expected of a fully turbulent boundary layer under the same pressure gradient, which resembles more closely the RANS prediction with a skin friction peak much closer to the end of the acceleration. 
Therefore, figure~\ref{fig:CpCf} suggests that the near-wall region of the boundary layer over the bump is indeed experiencing relaminarization.

\begin{figure}
\centering
  \includegraphics[width = 0.97\textwidth]{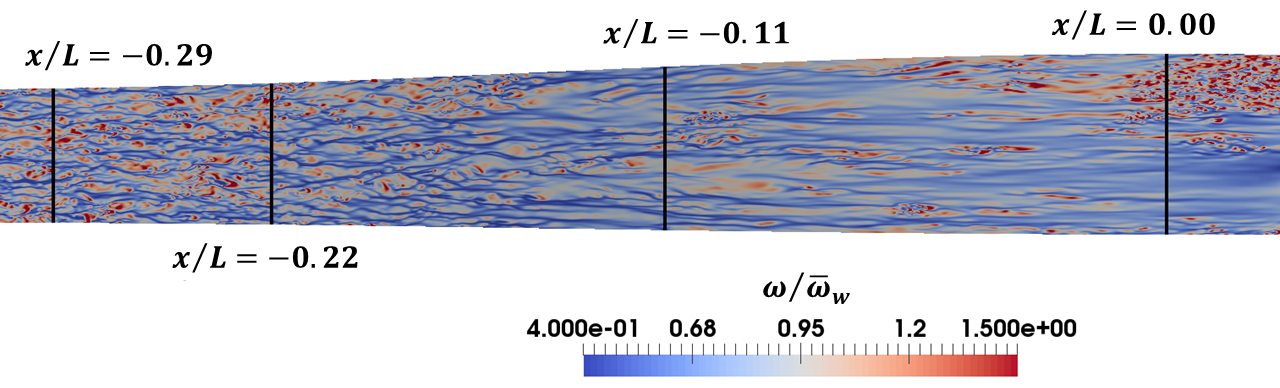}
  \includegraphics[width = 0.97\textwidth]{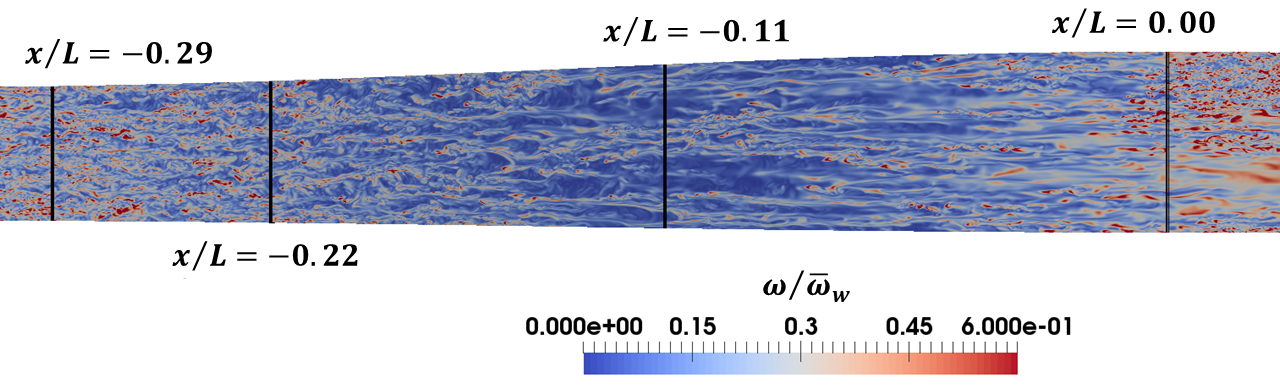}
  \includegraphics[width = 0.97\textwidth]{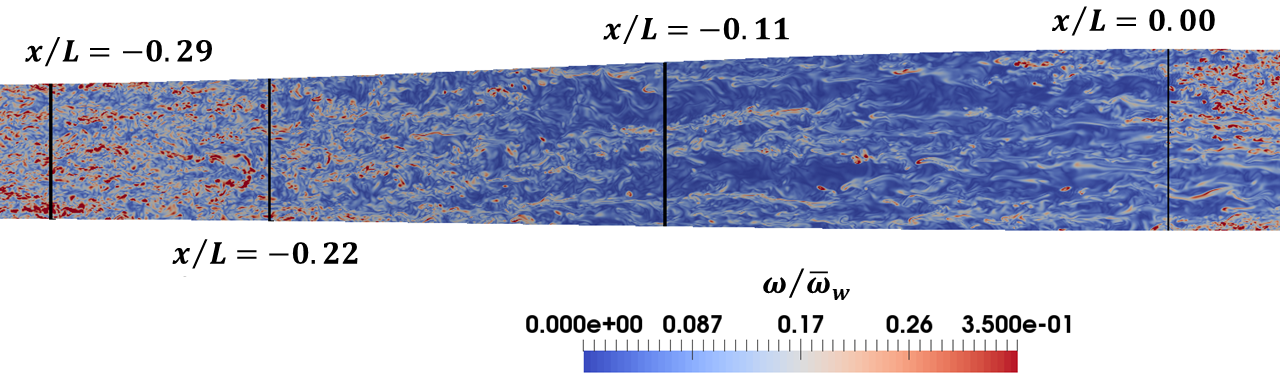}
  \includegraphics[width = 0.97\textwidth]{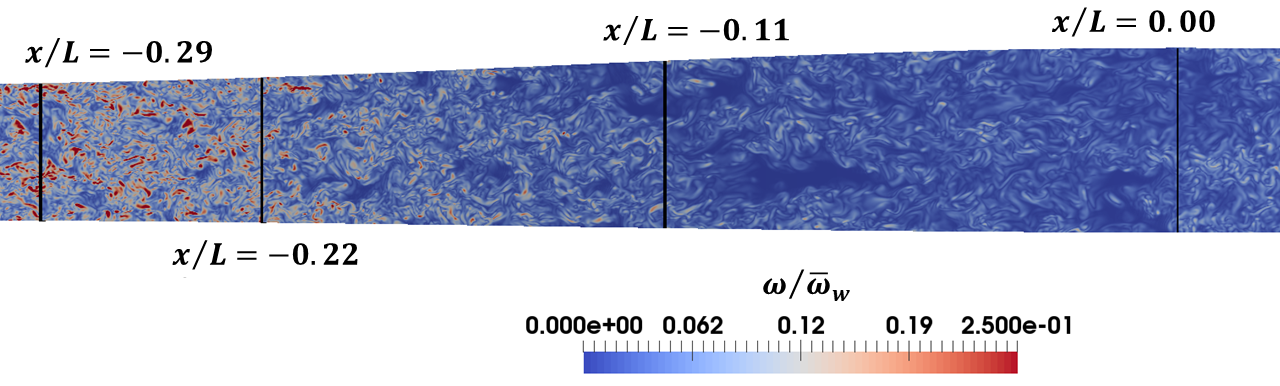}
  \caption{Instantaneous vorticity magnitude normalized by the local time- and spanwise-averaged wall vorticity at different locations within the boundary layer in the FPG region of the bump. From top to bottom, the heights above the wall of the slices are $n/\tilde{\delta}_{995}=0.01, 0.05, 0.1, 0.4$ or $n^+ \approx 7, 40, 75, 310$. The black vertical lines mark the location of key events of the flow.}
\label{fig:vortFPG}
\end{figure}

% Intermittency
Further credence to this claim comes by looking at instantaneous contours of vorticity at different heights within the boundary layer. Figure~\ref{fig:vortFPG} shows slices of the flow at a few locations of constant $n/\tilde{\delta}_{995}$. Note that because $\tilde{\delta}_{995}$ varies with downstream position, these surfaces track local boundary layer height, not distance to wall. The vorticity magnitude is normalized by the local (same streamwise location) time- and spanwise-averaged wall vorticity magnitude $\overline{\omega}_{w}$ in order to remove the increase of vorticity due to flow acceleration and highlight the fluctuations relative to their local mean wall value. Moreover, solid black lines across the width are placed to mark important landmarks within the FPG region. These are the start of the FPG at $x/L=-0.29$, the location where the streamwise velocity departs above the logarithmic law at $x/L=-0.22$ (discussed in more detail later), the peak strength of the FPG at $x/L=-0.11$, and the bump peak and end of the FPG at $x/L=0.00$.
Note that the same figures were also analyzed at different time steps in order to confirm that the following features are always present and not just at the time instant shown here.

Very close to the wall, the typical streamwise streaks of high and low vorticity are visible at the beginning of the FPG region. These remain fairly unchanged until around $x/L=-0.22$, downstream of which the streaks appear to stretch in the streamwise direction and reduce in relative intensity (recall the contours are normalized by the local mean wall vorticity). As the location of peak FPG strength is approached, the character of the turbulence is significantly altered. The streaks become very elongated in the streamwise direction while also thickening in the spanwise direction, and their relative intensity is significantly reduced. These changes continue past the peak FPG until the peak of the bump, where quiet flow with weak fluctuations surround a few regions with smaller and more intense scales. At the bump peak, larger regions of high turbulent intensity develop. These are signs of retransition and are discussed in greater detail in \textsection~\ref{sec:ResAPG}.

Slightly further from the wall at $n/\tilde{\delta}_{995}=0.05$ and $0.1$, similar trends can be observed. Fully turbulent flow is present at the start of the FPG and is maintained until slightly downstream of $x/L=-0.22$, after which the turbulent scales weaken relative to the wall vorticity, stretch significantly in the streamwise direction, and quiet regions of low vorticity appear. At this height above the wall, the streaks of high vorticity fluctuations turn into very thin and long ridges, while the quiet areas form ``valleys'' that are much wider but equally as long.
These are clear signs that the near-wall region of the boundary layer is no longer fully turbulent, and instead becomes intermittent and approaches a quasi-laminar state. A relaminarization process is therefore taking place in this flow. The contours also show how this truly is a gradual process since no clear demarcation can be identified from these instantaneous fields across which the turbulence suddenly changes from fully turbulent to intermittent. Moreover, for this Gaussian bump the onset of relaminarization occurs upstream of the peak strength of the FPG, where the pressure gradient parameters $K$ and $\Delta_p$ are even further from the critical values reported in other studies (see figure~\ref{fig:pGrad}). 

In the outer layer at $n/\tilde{\delta}_{995}=0.4$, the turbulence is also very much affected by the strong FPG. As the flow is being accelerated, the normalized vorticity fluctuations weaken with the smaller scales decaying entirely while the larger scales remain, which is consistent with the decrease in Reynolds number $Re_{\tilde{\theta}}$ throughout the FPG. Contrarily to the near-wall region, however, the shape of the fluctuations does not appear to be significantly impacted.

As was noted earlier, both concave and convex streamline curvature effects are acting during the FPG, with the change from the former to the latter being located at $x/L=-0.14$. The vorticity contours in figure~\ref{fig:vortFPG} clearly show that changes to the turbulence are taking place where the destabilizing concave curvature is acting, well before the stabilizing convex curvature takes effect. The pressure gradient effects are therefore dominating over the curvature during most of FPG and the onset of relaminarization is attributed to the acceleration alone, rather than to the combination of the two forces.

% Retransition
Finally, the onset of relaminarization can be implied by evidence of retransition (or a return to fully turbulent flow) where the strong FPG is relaxed and the APG starts. The details of this process are discussed later in \textsection~\ref{sec:ResAPG}, however the combined evidence of the sudden increase in skin friction in figure~\ref{fig:CpCf}, the appearance of turbulent spots in the near-wall region in figures~\ref{fig:wallVort},~\ref{fig:vortFPG},~\ref{fig:vortAPG}, and 
the sudden enhancement of the Reynolds stresses in figure~\ref{fig:stressesAPG} further support the statement that the upstream boundary layer does not remain fully turbulent and instead experiences the effects of a relaminarization process to a moderate extent.

% summary
In spite of the guidelines offered by previous studies regarding various pressure gradient parameters and integral quantities, it is therefore determined that the boundary layer over the Gaussian bump at this Reynolds number does indeed enter a relaminarization process. Relaminarization is the process by which a fully turbulent boundary layer progresses towards a quasi-laminar state and fully turbulent flow is clearly no longer present in the near-wall region of this flow under the effects of the strong acceleration. 
The process, however, does not reach completion and, while the inner layer turbulence becomes intermittent, a quasi-laminar state is never achieved.
The disagreement with previous studies is likely caused by the added complexities of this flow. The literature mainly focused on equilibrium sink flows or boundary layers over flat plates that were monotonically accelerated from a zero pressure gradient by variation of the top boundary condition. The boundary layer over the Gaussian bump is deeply different; it experiences an APG upstream of the acceleration which, while being mild, affects the velocity and stress profiles in a significant manner, there is a high degree of non-equilibrium in the sense of the Clauser pressure gradient parameter, streamline curvature effects are present throughout the entire flow, and the Reynolds number is lower than what most experimental studies were able to achieve.
Furthermore, given the results discussed above, a definite marker or critical value marking the onset of relaminarization was not found. This is not a surprising conclusion, and in fact agrees with the interpretation offered by \citet{Sreenivasan82} and seconded by \citet{Warnack1998} that it is an ill-defined exercise to do so since relaminarization is a gradual process and not a catastrophic event. By extension, this study provides further evidence that critical parameters of any specific quantity should only be used as guidelines to suggest that reverse transition might be occurring, but further analysis of the state of the turbulence is needed for a more definite statement to be made. 
Moreover, a turbulence model based on these parameters is likely to not be universal.

\begin{figure}
\centering
  \includegraphics[width = 0.97\textwidth]{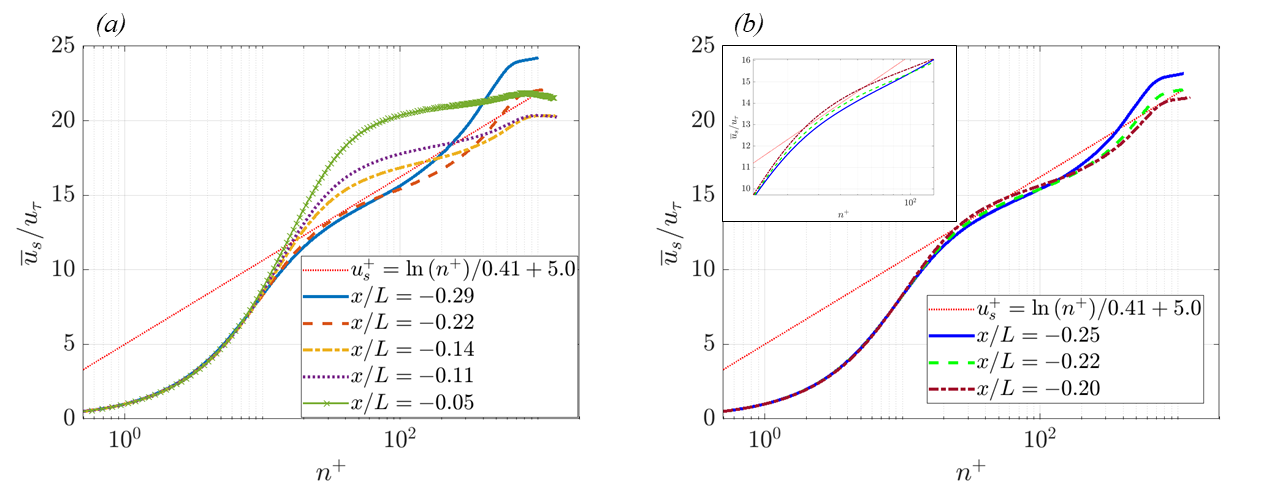}
  \caption{Mean streamwise velocity profiles in the FPG region normalized by wall units.}
\label{fig:velFPG}
\end{figure}

Figure~\ref{fig:velFPG}(a) shows the non-dimensional streamwise velocity profile at a number of locations on the upstream side of the bump.  
At the start of the FPG ($x/L=-0.29$), the velocity is consistent with the discussion in the previous section regarding the mild APG. The velocity has shifted slightly below the standard log law ($\kappa=0.41$ and $B=5.0$) but a logarithmic section remains and the wake is increased. Once again, the boundary layer entering the strong FPG is not the canonical ZPG flow.
Soon after that at $x/L=-0.22$, the FPG effects become clear. The wake is significantly reduced and the profile lacks a linear region that would result from a logarithmic relationship of the form $u_s^+=\ln{(n^+)}/\kappa+B$. There is, therefore, the breakdown of the logarithmic law with the velocity remaining for the most part below the law.
As the flow progresses through the FPG, the effects become increasingly stronger. The velocity continues to rise above the log law until the peak of the bump, with a thickening of the viscous sublayer and a continuous reduction of the wake in the outer layer. Note that convex curvature effects are also present downstream of $x/L=-0.14$, however the continued reduction of the wake downstream of this location indicate that the pressure gradient is dominating over curvature on the velocity. This result is consistent with the experimental study of \citet{Schwarz1996}.

Additionally, figure~\ref{fig:velFPG}(b) shows the velocity profiles at three locations in the proximity of the critical parameter $\Delta_\tau=-0.013$ where the initial departure above the logarithmic law has been identified \citep{Bradshaw1969}. From figure~\ref{fig:pGrad}, this critical parameter is observed at $x/L=-0.22$. Slightly upstream of this location, the velocity is entirely below the log law in the range $10 \le n^+ \le 100$, however slightly downstream it rises above for $25 \le n^+ \le 60$. At the location of $\Delta_\tau=-0.013$, the velocity osculates the law at $n^+=35$ and is below otherwise. 
These results are in great agreement with \citet{Bradshaw1969} and the notion of \citet{Patel_ReverseTrans_1968} that the non-dimensional shear stress gradient is a suitable quantity to measure the departure from the logarithmic law.
Furthermore, following the description of strong FPG flows in \citet{Sreenivasan82}, these effects are characteristic of the laminarescent boundary layer that may precede relaminarization. 
Regardless of the nomenclature used, when considered along with the non-dimensional vorticity contours in figure~\ref{fig:vortFPG}, the velocity profiles provide further evidence that: 1) strong FPG cause the breakdown of the standard logarithmic law and a departure above it while the flow is still fully turbulent, 2) the breakdown of these standard laws comes with a change in the fundamental character of the turbulence (significant changes in the vortical structures in figure~\ref{fig:vortFPG} can be seen to gradually occur across the $x/L=-0.22$ line at all values of $n/\tilde{\delta}_{995}$), and 3) these changes to the velocity occur at much smaller values of the non-dimensional pressure gradients relative to the critical ones ($K=1.4 \times 10^{-6}$ where $\Delta_p=\Delta_\tau=-0.013$) and further upstream than evidence of intermittency in the near-wall turbulence and thus these alone should not be considered a sign of the onset of relaminarization.

\begin{figure}
\centering
  \includegraphics[width = 0.97\textwidth]{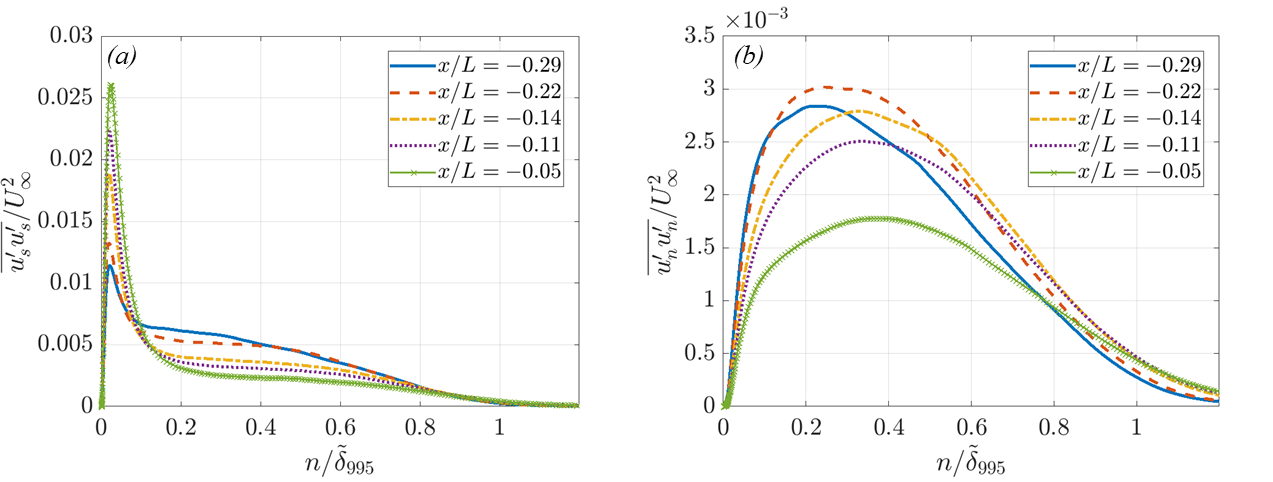}
  \includegraphics[width = 0.97\textwidth]{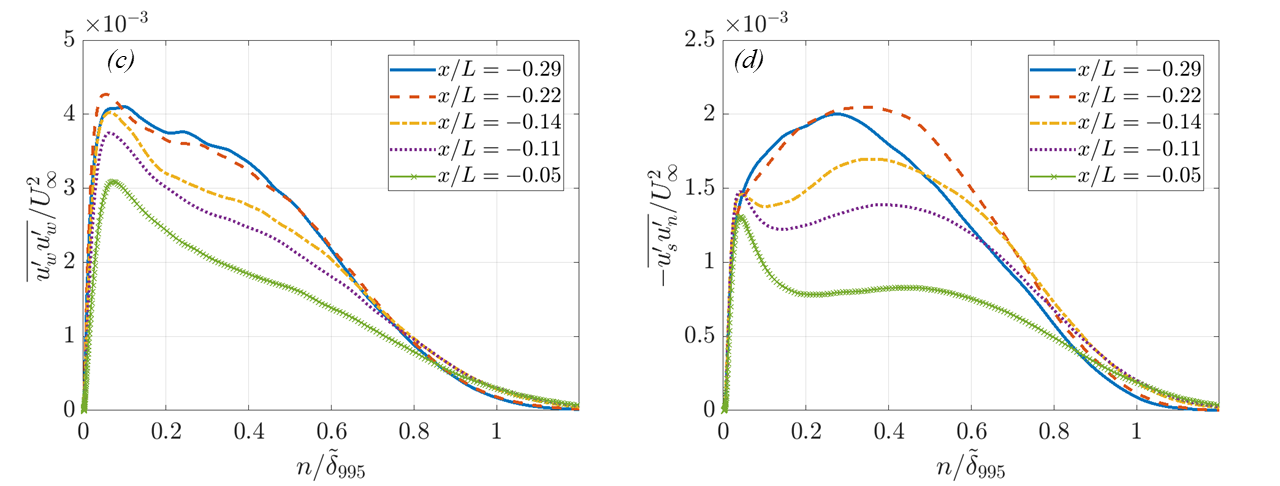}
  \includegraphics[width = 0.97\textwidth]{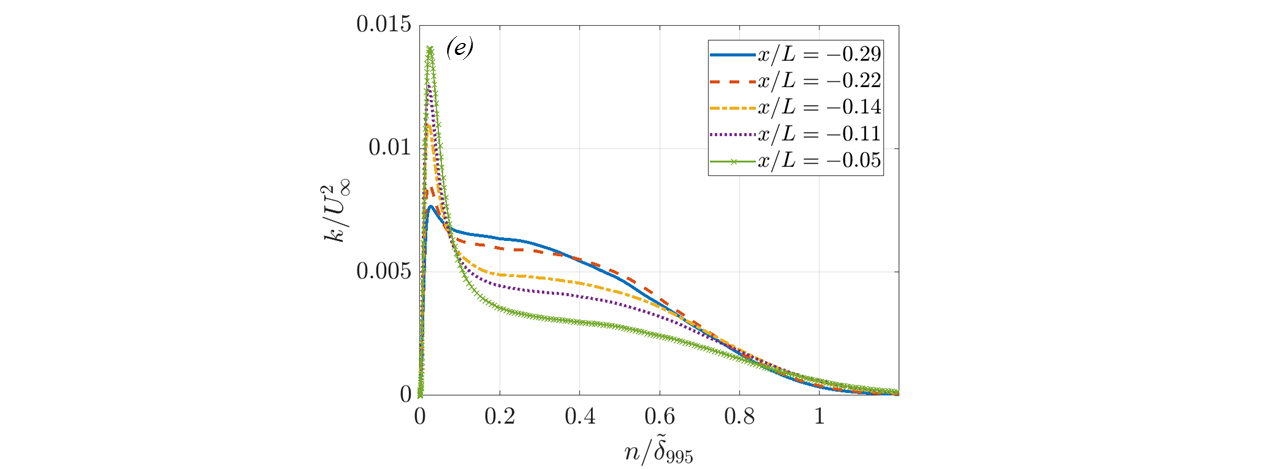}
  \caption{Mean Reynolds stresses and turbulent kinetic energy in the FPG region of the Gaussian bump normalized by $U_\infty$.}
\label{fig:stressesFPG}
\end{figure}

\begin{figure}
\centering
  \includegraphics[width = 0.97\textwidth]{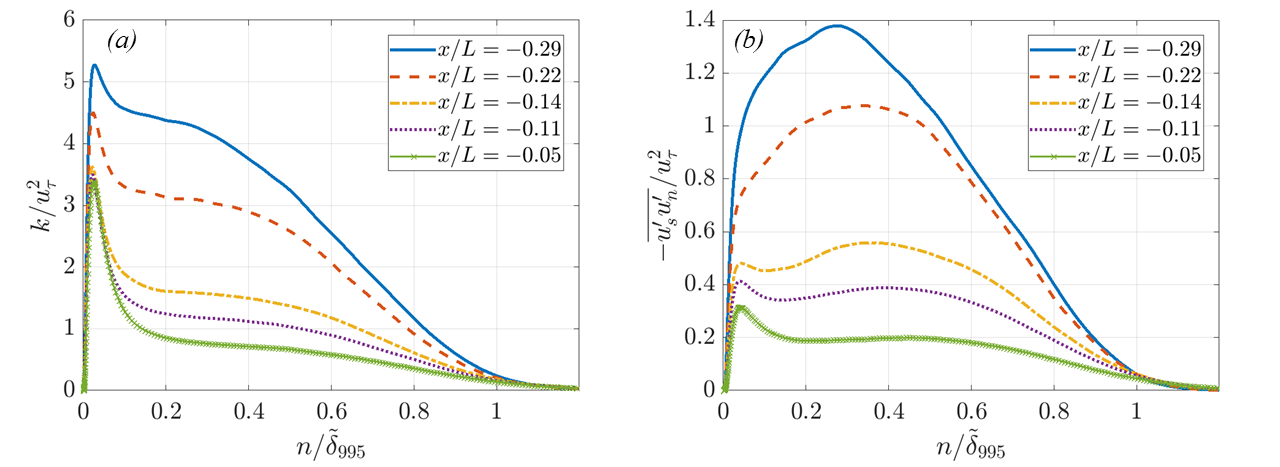}
  \includegraphics[width = 0.97\textwidth]{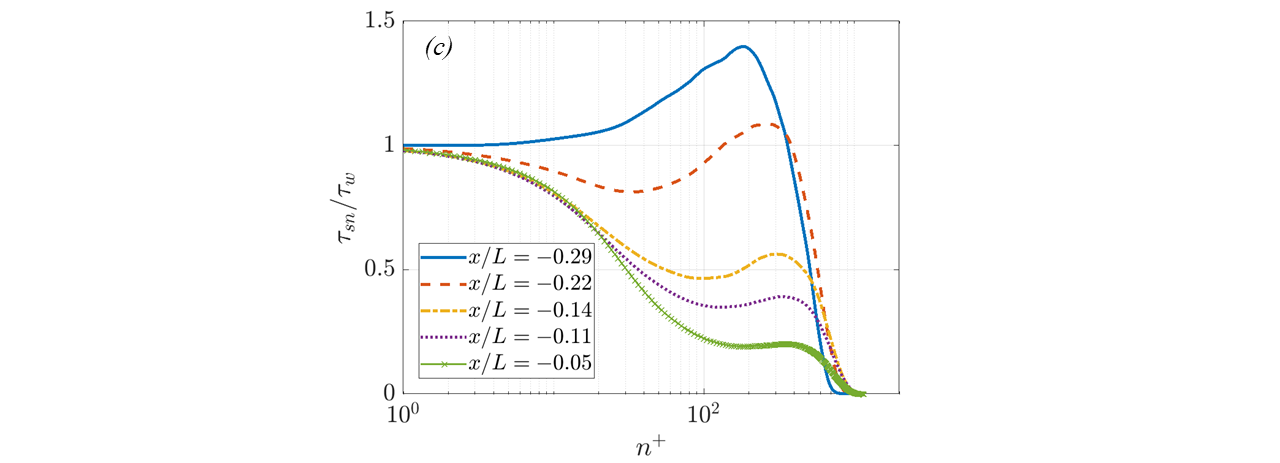}
  \caption{Mean turbulent kinetic energy (a), Reynolds shear stress (b), and total shear stress (c) in the FPG region of the Gaussian bump normalized by wall units.}
\label{fig:stressesPlusFPG}
\end{figure}

% Reynolds Stresses
Reynolds stress and turbulent kinetic energy profiles in the FPG region are shown in figure~\ref{fig:stressesFPG} normalized by the freestream velocity $U_\infty$. The same streamwise stations as those in figure~\ref{fig:velFPG} are presented. 
%Recall that convex curvature of the surface starts at $x/L=-0.14$, therefore in the initial part of the FPG the mean streamlines are concave.
The streamwise fluctuations develop a large inner peak and an outer knee point, which progressively increases and decreases, respectively, as the turbulence advects through the pressure gradient. The TKE, which is heavily dependent on $\overline{u^{\prime^{2}}_s}$, follows very similar trends. The increase of the streamwise Reynolds stress near the wall is consistent with the instantaneous contours of vorticity and the wall streaks observed in figures~\ref{fig:wallVort} and~\ref{fig:vortFPG}. 
The wall-normal fluctuations $\overline{u^{\prime^{2}}_n}$ show an increase in intensity in the initial part of the FPG before the deviation above the logarithmic law, however after this event a significant reduction in the peak value is observed. The peak value moves further from the wall, further reducing this component of the stress near the wall.
The spanwise Reynolds stress behaves similarly, with only a slight change at $x/L=-0.22$, followed by a decrease further downstream. The shape does not appear to be altered in a major way, rather the magnitude is scaled down. 
The fact that all three normal components of the Reynolds stresses decrease in the outer layer, resulting in the reduction of the TKE, is consistent with the slice at $n/\tilde{\delta}_{995}=0.4$ in Fig~\ref{fig:vortFPG} showing no significant alteration in the shape of the turbulent eddies but a reduction in intensity. 
The shear stress $-\overline{u'_su'_n}$ exhibits interesting features. At $x/L=-0.22$, the outer layer shows an increase in the stress, however, near the wall, the profile shows a sudden break (more visible in figure~\ref{fig:stressesPlusFPG} where the profiles are not as close together), suggesting a change in the near-wall turbulence. Downstream of this location, the profiles develop a bi-modal shape, with an inner peak which appears to be relatively fixed in magnitude and distance from the wall ($n/\tilde{\delta}_{995} = 0.04$), and an outer peak which continuously decreases and moves further from the wall. The inner peak location also matches the break in the profile at $x/L=-0.22$.

The appearance of knee points and multiple peaks in the Reynolds stress profiles is indicative of an internal layer being present in the flow. This is a common feature of boundary layers over bumps and hills. Previous studies identified the cause of these internal layers to be due to a sudden change in the wall boundary conditions and thus shear stress, including changes in pressure gradients \citep{Tsuji_hills} and surface curvature discontinuities \citep{Baskaran_part1,Webster_bump,Wu_LESbump}. 
Since the Gaussian shape of the bump ensures continuity of the curvature, the internal layer observed here is caused by the change in pressure gradient from adverse to favorable at $x/L=-0.29$. 
Moreover, in the presence of internal layers, the inner and outer layers become almost independent of each other, with the latter behaving similarly to a free-shear layer \citep{Baskaran_part1}. 
The internal layer has a constant Reynolds shear stress throughout the FPG as seen by the effectively constant magnitude of the inner peak in figure~\ref{fig:stressesFPG}, which is a feature which was noted for other strongly accelerated flows \citep{Narasimha_Relaminarization_1979}, and a rapidly evolving anisotropy of the normal stresses in favor of the streamwise direction and at the expense of the wall-normal direction predominantly. This behavior is consistent with the instantaneous contours of vorticity in figure~\ref{fig:vortFPG}.
Moreover, a closer look along the initial part of the FPG reveals that the break in the shear stress profile near the wall appears as soon as the pressure gradient changes sign. Similarly, the near-wall peak of the TKE starts to grow immediately downstream of $x/L=-0.29$. Since the curvature changes direction approximately half way along the FPG, these two inner layer effects appear to be strongly related to the pressure gradient rather than the curvature.
In the outer layer, the behavior of the Reynolds stresses is slightly more complex. 
The TKE is reduced from the start of the FPG and develops a knee point rapidly. 
On the contrary, the shear stress is observed to gradually increase in the outer layer relative to the profile at $x/L=-0.29$ until around $x/L=-0.24$ and then decrease as shown in figure~\ref{fig:stressesFPG}. A similar response is observed for the wall-normal and spanwise components as well. Interestingly, $x/L=-0.24$ is the location where the parameter $\hat{\kappa}$ indicates the concave curvature effects are a maximum upstream of the bump. This points to the fact that the outer layer shear stress is initially responding to the curvature effects, which at this location dominate over the FPG effects. Eventually, as the concavity is reduced and the strength of the FPG increases, pressure effects dominate and the Reynolds stresses in the outer layer are reduced. Downstream of $x/L=-0.14$, the stabilizing effects of the convex curvature are compounded with the FPG ones resulting in a rapid decrease of the shear in the outer layer. 

This analysis, particularly of the shear stress, strongly suggests the independence of the inner and outer layers. The former is dominated by favorable pressure gradient effects, while the latter responds to a combination of streamline curvature and pressure gradient. It is the inner layer physics, however, that are mostly responsible for the skin friction coefficient over the bump. Consequently, a turbulence model that hopes to be predictive of a flow of this kind must be able to capture these near-wall effects. Note that the Gaussian bump flow considered here is not an isolated test case of these effects, and similar examples exist in the literature \citep{Wu_LESbump,Uzun_NASAHumpLES_2018,Matai_BumpLES_2019}.

Further interesting observations arise from the TKE and Reynolds shear stress profiles normalized by wall units in figure~\ref{fig:stressesPlusFPG}. This non-dimensionalization is useful to show changes in these quantities relative to the local wall shear rather than a comparison of their absolute magnitudes. Additionally, note that $\tau_{sn}$ represents the total shear stress, which is the sum of the Reynolds and viscous stresses.
As the turbulence advects through the FPG, the wall-shear normalized TKE is significantly reduced everywhere across the boundary layer, including at the peak. This trend of the near-wall TKE is in contrast to the one shown in figure~\ref{fig:stressesFPG}, and indicates that while the streamwise fluctuations are strengthening near the wall as the flow is being accelerated, they are not increasing fast enough relative to the increase in wall shear stress. Moreover, the drop in TKE in the outer layer is magnified with this non-dimensionalization. These trends are in full agreement with the vorticity contours in Fig~\ref{fig:vortFPG} given the similar non-dimensionalization used.
The Reynolds shear stress exhibits similar trends to the TKE showing significant reduction due to the FPG. In this non-dimensional form, the inner peak is not constant and instead continues to diminish reaching values as low as 35\% of the wall shear stress (a density of one was used in the simulation resulting in $u_\tau^2=\tau_w$).
The non-dimensional total shear stress in figure~\ref{fig:stressesPlusFPG} clearly shows the significant drop in turbulent stress relative to the viscous stress both in the inner and outer layers. Considering both the Reynolds and total shear stress profiles in the figure, at $x/L=-0.05$ the viscous stress dominates over the turbulent one until about $n^+=50$, clearly pointing to a thickening of the viscous sublayer measured in wall units.
This in turn leads to a very steep wall-normal gradient at a distance between $10 \le n^+ \le 60$, which directly impacts the streamwise momentum equation and thus the streamwise velocity as seen in figure~\ref{fig:velFPG}. In other words, it is the steep gradient in the total shear stress that is responsible for the deviation above the logarithmic law. Once this gradient reaches a certain magnitude, measured in part by $\Delta_\tau$, the breakdown of the logarithmic law is to be expected.
Note that while the relative size of the Reynolds shear stress is significantly reduced by the FPG, it is not negligible and thus the relaminarization process does not reach completion and the flow is still turbulent, albeit only intermittently in the near-wall region. The outer layer, by contrast, remains fully turbulent with the fluctuations decaying in intensity.
%The shear stress gradient near the wall is largely dependent on the pressure gradient, however the two are not the same (see figure~\ref{fig:pGrad} although one could imagine other changes to the wall boundary conditions that would cause an even larger discrepancy), therefore making $\Delta_\tau$ a better measure of this phenomenon than $\Delta_p$. Non-dimensionalization by the local $u_\tau$, therefore, brings to light the significant weakening of the turbulence by the FPG and curvature effects. The turbulent kinetic energy and shear stress are not keeping up with the increase in wall shear due to the flow acceleration.

\begin{figure}
\centering
  \includegraphics[width = 0.55\textwidth]{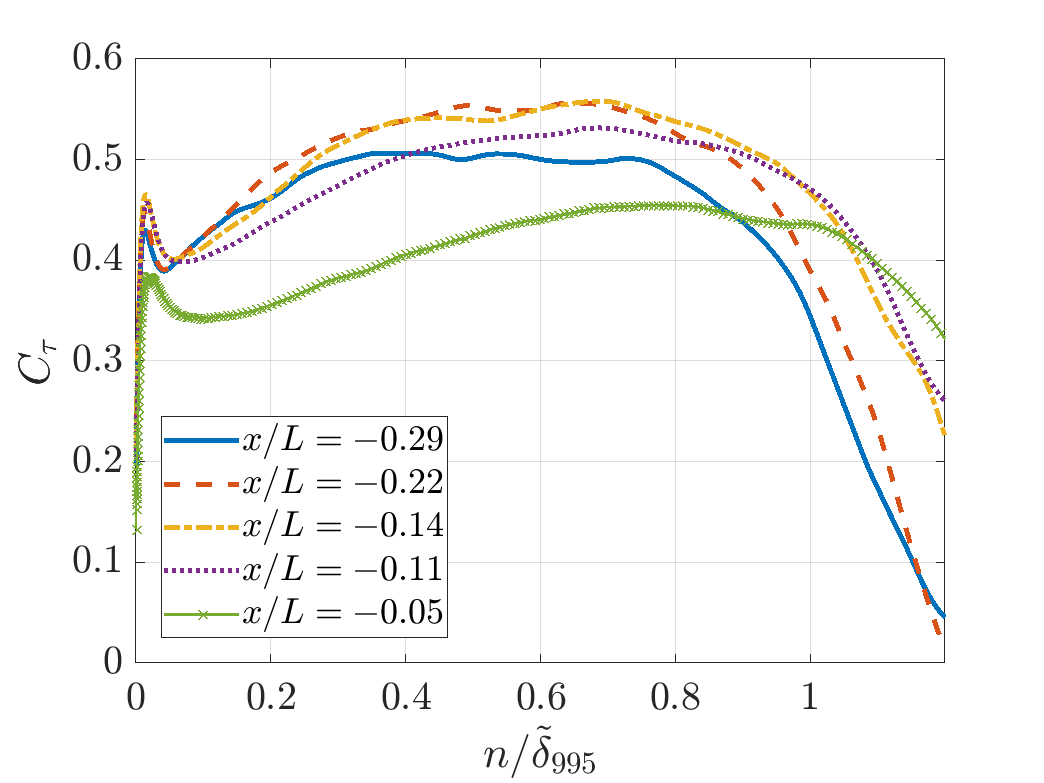}
  \caption{Correlation coefficient $C_\tau$ in the FPG region of the Gaussian bump.}
\label{fig:CtauFPG}
\end{figure}

% Correlation
Finally, the correlation coefficient defined as 
\begin{equation}
C_\tau=\frac{-\overline{u'_su'_n}}{\sqrt{\overline{u_s^{\prime^{2}}}} \sqrt{\overline{u_n^{\prime^{2}}}}}
\label{eq:Ctau}
\end{equation}
is plotted in figure~\ref{fig:CtauFPG} for the usual streamwise locations along the FPG.
At $x/L=-0.29$, the coefficient has the canonical value of 0.5 in the outer layer, and slightly lower near the wall. Further downstream in the region $-0.29 < x/L \le -0.14$, the streamwise and wall-normal fluctuations become slightly more negatively correlated (note the negative sign in \eqref{eq:Ctau}) everywhere within the BL. Given the strength of the pressure gradient in this region, this change is attributed to the concave curvature of the mean streamlines which is a maximum at $x/L=-0.24$. Furthermore, from the vorticity contours in figure~\ref{fig:vortFPG} it is evident that the character of the turbulence is significantly altered by $x/L=-0.14$, yet the profile of $C_\tau$ at this location shows only a slight change. 
Only downstream of $x/L=-0.14$, where the curvature changes to convex, $C_\tau$ decreases faster as the boundary layer approaches the bump peak where $\hat{\kappa}$ is a local maximum. 
These results are in agreement with \citet{Narasimha_Relaminarization_1979,Spa86,SoMellor_Convex_73}, and support other observations that the strong FPG does not change the correlation between these two fluctuations of the turbulence even though the Reynolds stresses are significantly reduced and the turbulence character is significantly altered during relaminarization.

% Summary

% APG
\subsection{Bump Peak and Strong Adverse Pressure Gradient}
\label{sec:ResAPG}
% Discuss APG effects on velocity and stresses

As shown by the skin friction coefficient in figure~\ref{fig:CpCf} and the contours of instantaneous vorticity in figures~\ref{fig:wallVort} and~\ref{fig:vortFPG}, in the vicinity of the bump peak where the FPG is relaxed and pressure gradient changes from favorable to adverse there is a sudden enhancement of the near-wall vorticity which leads to a rise in the wall shear. This feature is of course related to the significant weakening of the near-wall turbulence caused by the strong upstream FPG and the onset of relaminarization. It is, in fact, a partial retransition to fully turbulent flow, where the word ``partial'' is used since the upstream flow did not reach the quasi-laminar state associated with the completion of the relaminarization process. In this section, the details of the flow and turbulence as they move through this segment of the bump are discussed.

\begin{figure}
\centering
  \includegraphics[width = 0.97\textwidth]{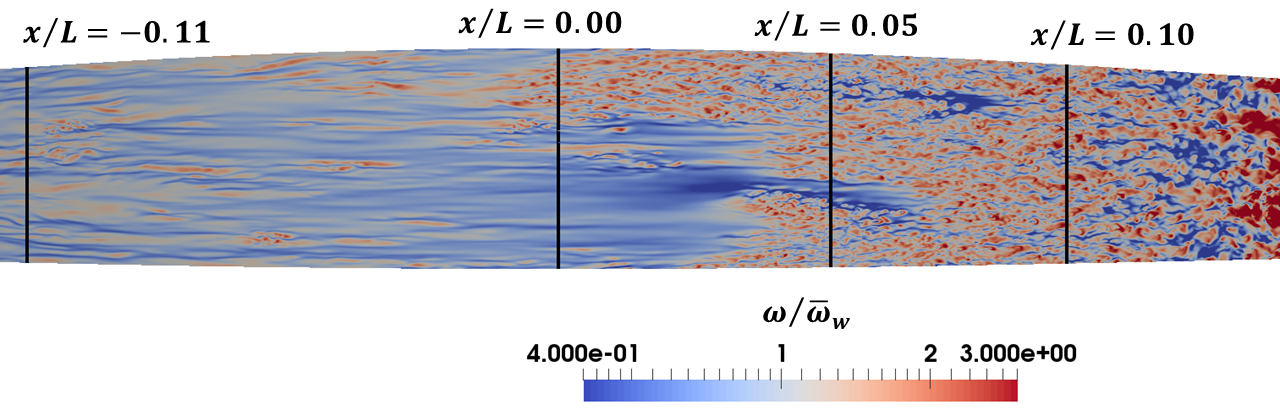}
  \includegraphics[width = 0.97\textwidth]{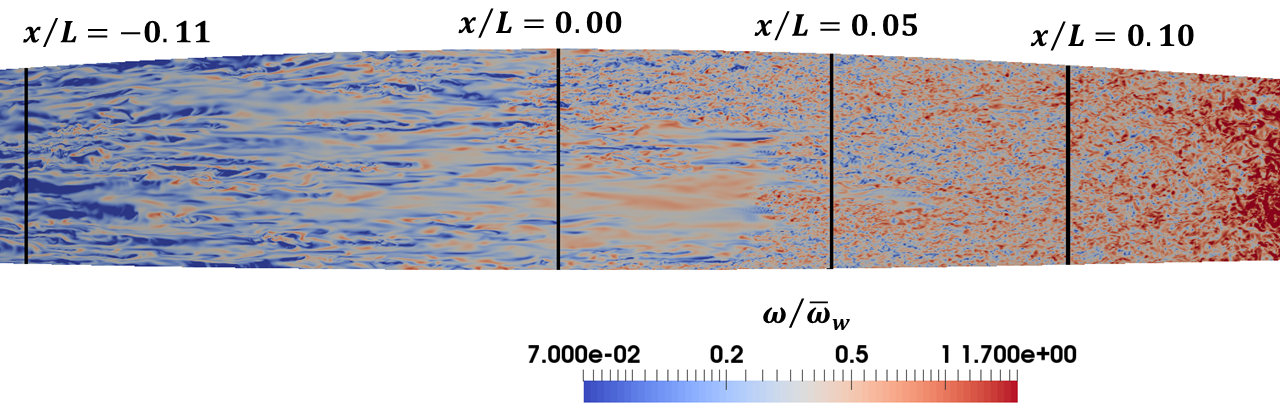}
  \includegraphics[width = 0.97\textwidth]{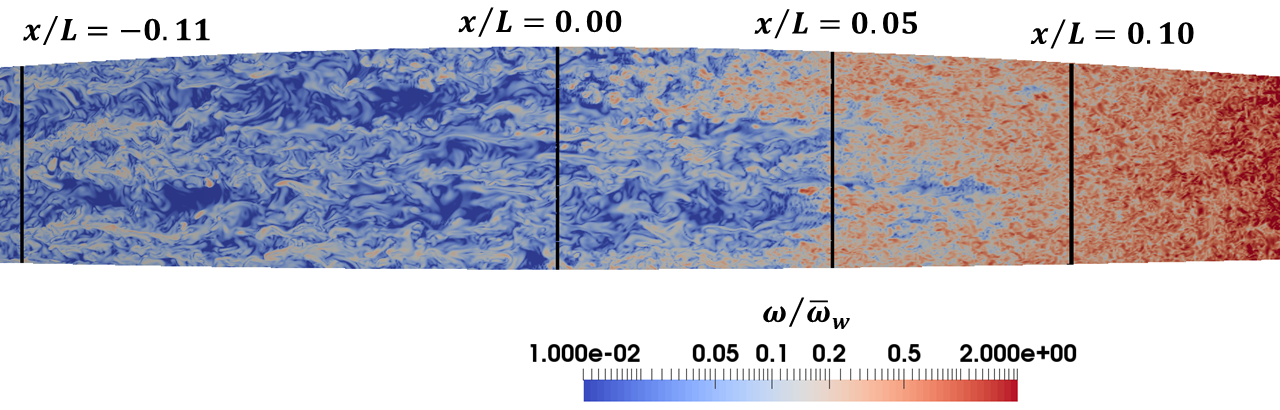}
  \includegraphics[width = 0.97\textwidth]{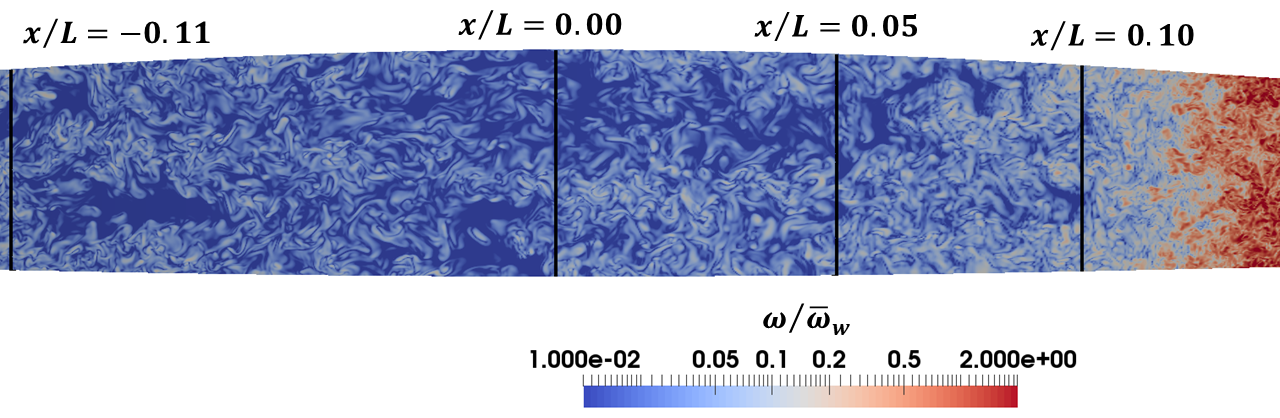}
  \caption{Instantaneous vorticity magnitude normalized by the local time- and spanwise-averaged wall vorticity at different locations within the boundary layer in the vicinity of the bump peak. From top to bottom, the heights above the wall of the slices are $n^+=5, 30, 100, 300$. The black vertical lines mark the location of key events of the flow.}
\label{fig:vortAPG}
\end{figure}

% Vorticity
Figure~\ref{fig:vortAPG} shows slices of instantaneous vorticity magnitude normalized by the local time- and spanwise-averaged vorticity at the wall over the bump peak. The slices are taken at different heights within the boundary layer, in this case measured in wall units with $n^+$. Note that streamwise changes in mean $u_\tau$ are accounted for, thus the slice is at the same height above the wall in local wall units but not in physical units since $u_\tau$ is not constant. Moreover, black lines across the domain are used to mark the location of key events of the boundary layer flow. These are the peak strength of the FPG at $x/L=-0.11$, the bump peak and change in sign of the pressure gradient at $x/L=0.00$, the small local maximum in $C_f$ at $x/L=0.05$, and approximately half way between the peak and incipient separation at $x/L=0.10$. Finally, a logarithmic scale is used in order to highlight turbulent structures with values of the normalized vorticity differing by orders of magnitude. 

Starting with the slice closest to the wall at $n^+=5$, turbulent spots, which may be identified as clumps of small-scale turbulent structures of high intensity, are seen to form as early as upstream of the bump peak where the FPG is relaxed. These are intermittent with regions of very quiet and weak vorticity fluctuations, and appear to grow in size and intensity as they progress downstream into the strong APG. As noted by other studies on relaminarizing boundary layers \citep{Narasimha1973,Blackwelder72,Sreenivasan82}, these are clear signs of a retransition process taking place near the wall. As the FPG is relaxed, the stabilizing effect of the acceleration diminishes and instabilities are allowed to grow once again to produce a new fully turbulent internal layer. The presence of fluctuations of different scales in the incoming flow makes this process often very sudden and energetic, as is evident in figure~\ref{fig:vortAPG}.
It is important to note that this phenomenon is a feature of the upstream strong FPG and relaminarization process, and is not due to the APG. In fact, previous studies exhibited this phenomenon even with the strong FPG relaxing into a ZPG region. Nevertheless, the destabilizing effects of the APG certainly aid in the onset of instability and growth of the turbulent spots, accelerating the formation of fully turbulent flow.
At $x/L=0.05$, just four local boundary layer thicknesses downstream of the bump peak, the boundary layer at this height above the wall is effectively fully turbulent. The scales at this location are small and fairly isotropic in shape, but grow in size and stretch in the streamwise direction by $x/L=0.10$. The structure of the canonical boundary layer (see figure~\ref{fig:vortFPG} at the start of the FPG) is not recovered however.

\begin{figure}
\centering
  \includegraphics[width = 0.55\textwidth]{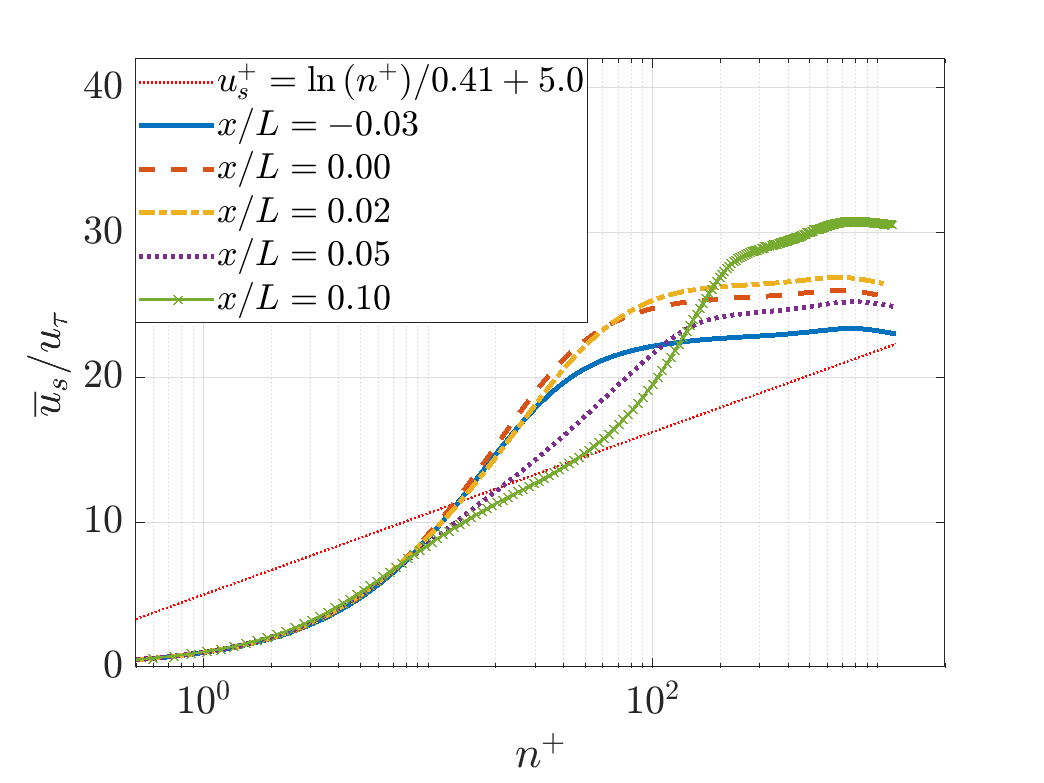}
  \caption{Mean streamwise velocity profiles in the APG region normalized by wall units.}
\label{fig:velAPG}
\end{figure}

The other slices in figure~\ref{fig:vortAPG} show the height of the turbulent spots and their growth rate in the wall-normal direction. In their initial stages, the turbulent spots extend well past $n^+=30$ but are barely visible at $n^+=100$. Signs of the new internal layer only become visible further downstream at this height. 
This indicates clearly that this phenomenon originates near the wall and propagates away from it as the internal layer grows into the boundary layer.
The highest location above the wall, $n^+=300$, shows the flow in the outer wake of the boundary layer (see the velocity profiles in figure~\ref{fig:velAPG}) as it flows from the FPG into the APG. In this region, no change in the size, structure, and strength of the turbulent scales is visible from the contours of vorticity. The only change is observed downstream of $x/L=0.10$ when the internal layer finally reaches this height above the wall. This behavior is consistent with the notion that the inner and outer layers of this flow are almost independent of each other, with the latter behaving similarly to a free-shear layer subject to the strong convex curvature effects that are present around the peak of the bump. By contrast, the near-wall physics are dominated by the pressure gradient and their associated relaminarization and retransition that are not directly felt in the outer layer.

% Velocity
The mean streamwise velocity profiles are presented in figure~\ref{fig:velAPG} for a number of locations around the bump peak and the initial part of the APG region. Note that $x/L=0.02$ corresponds to the local minimum in the $C_f$ curve.
At the end of the FPG and at the bump peak, significant deviation above the logarithmic law is still visible. In fact, the deviation continues to grow throughout the entire FPG with the largest departure being at the peak and slightly downstream of it. At $x/L=0.02$ the velocity still exhibits the characteristics of the upstream FPG. This is consistent with the contours in figure~\ref{fig:vortAPG} since the flow is still heavily intermittent at this location.
Further downstream at $x/L=0.05$, the effects of the partial retransition become visible. The velocity gradient in the standard logarithmic region is reduced along with the thickness of the viscous sublayer due to the sudden surge in shear stress discussed later in this section.
At the last station plotted, the velocity approaches the standard log law even closer and appears to have two distinct regions. The first is found below $n^+=40$ and resembles the canonical turbulent boundary layer shape with a buffer layer and a semi-linear section (although with a different slope and intercept from the standard values). This is caused by a new internal layer forming at the bump peak which has reached heights around $n^+=100$ at this streamwise location.
The second region appears to be a very pronounced wake above the underlying internal layer which maintains the shape of the upstream FPG profiles. This is once again in agreement with the contours in figure~\ref{fig:vortAPG}.
Note that due to the strength of the APG and the imminent incipient separation of the flow at $x/L=0.19$, the velocity does not reach agreement with the standard logarithmic law before separation. This result is consistent with the vorticity contours at the wall in figure~\ref{fig:wallVort} which show that the standard wall streaks do not develop within the APG region.

\begin{figure}
\centering
  \includegraphics[width = 0.97\textwidth]{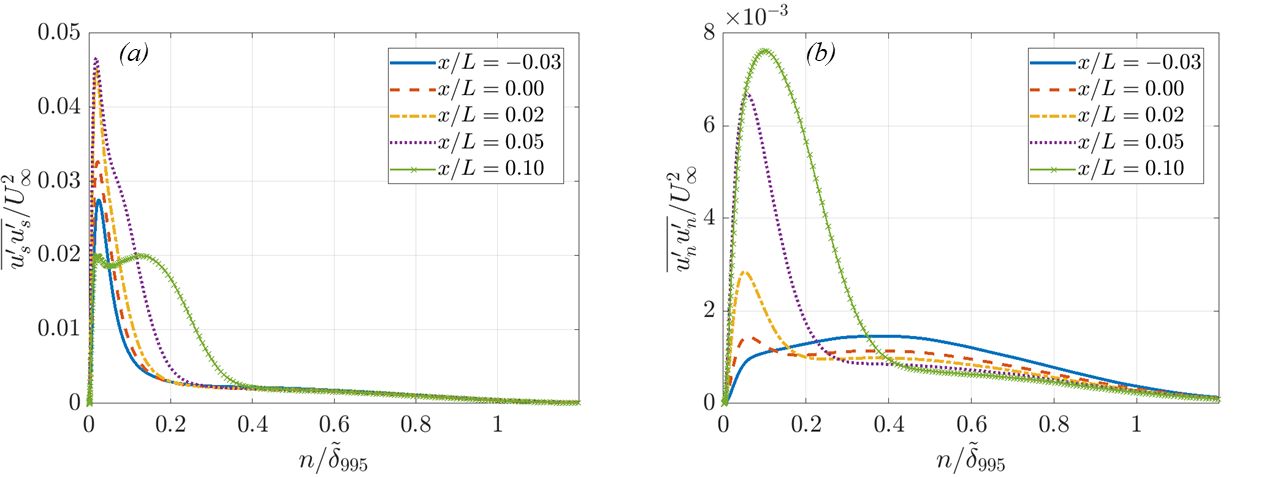}
  \includegraphics[width = 0.97\textwidth]{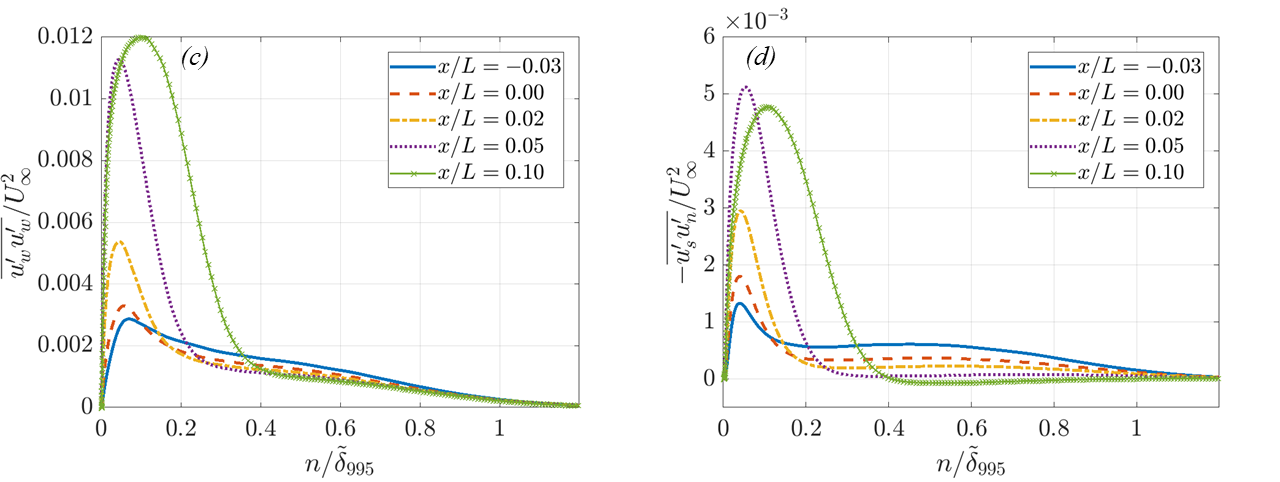}
  \includegraphics[width = 0.97\textwidth]{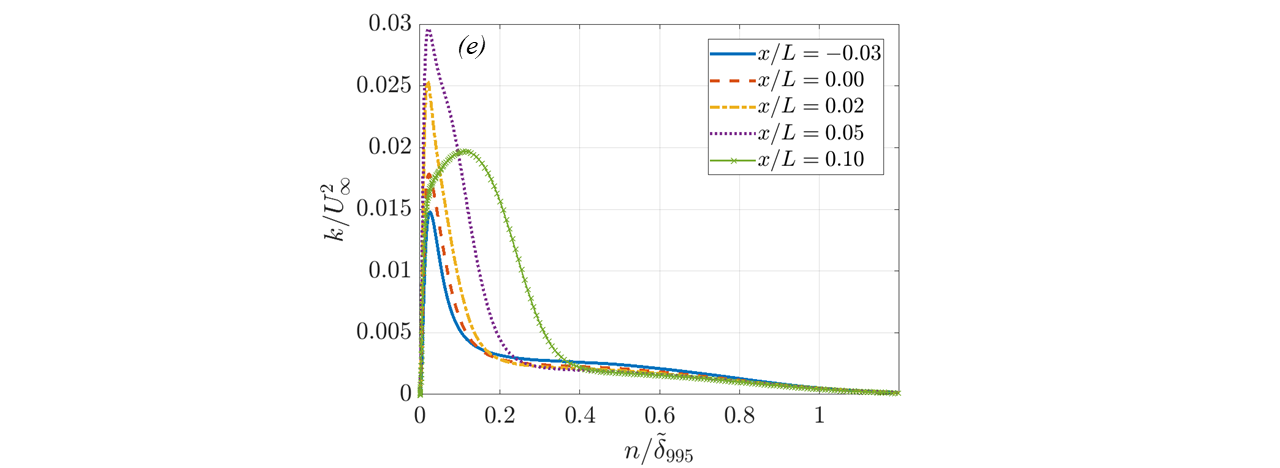}
  \caption{Mean Reynolds stresses and turbulent kinetic energy in the APG region of the Gaussian bump normalized by $U_\infty$.}
\label{fig:stressesAPG}
\end{figure}

% Stresses
The Reynolds stresses for the same locations are presented in figure~\ref{fig:stressesAPG}. 
Starting with the turbulent kinetic energy, at the end of the FPG and at the bump peak the profiles look similar to the ones in the upstream FPG, with a larger peak than the $x/L=-0.05$ location in figure~\ref{fig:stressesFPG}. This is consistent with the increase in the inner peak of this stress component due to the acceleration of the flow. There is also a further reduction in the TKE of the outer layer due to the continued FPG and convex curvature effects. Entering the initial part of the APG, a significant surge the value of the peak is observed, almost doubling at $x/L=0.05$ relative to the bump peak and then decreasing again after this station. This inner peak is also increasing in thickness with downstream location. These are clear signs of the new internal layer produced by the partial retransition. The outer layer TKE stays almost constant throughout the APG, providing further evidence of the separation of these two sections of the boundary layer. As the underlying internal layer grows in thickness, the outer layer is simply displaced further from the wall, resulting in a significant growth of the overall boundary layer as seen in figure~\ref{fig:BLdelta}.

The turbulent shear stress follows similar trends to the TKE, with some differences mainly in the outer layer. The inner peak also rises in magnitude and thickens significantly due to the new internal layer (more than a factor of three at $x/L=0.05$), however the outer layer shear is continuously reduced, eventually changing sign at $x/L=0.10$. Since the pressure gradient effects are now destabilizing, these are clearly convex curvature effects that are still present until $x/L=0.14$. Once again, the inner part of the boundary layer is responding to the pressure gradient, while the outer layer is responding to the mean streamline curvature.

All three normal Reynolds stresses generally behave as expected, increasing near the wall and staying fairly constant or slightly decreasing in the outer layer. They do react to retransition and the adverse pressure gradient slightly differently, however. The streamwise fluctuations respond the quickest, with the largest peak in figure~\ref{fig:stressesAPG} actually occurring at $x/L=0.02$. They also develop a tri-modal shape at $x/L=0.10$ that is not seen in other components of the Reynolds stress tensor. 
The wall-normal and spanwise fluctuations respond later but more drastically (the wall-normal fluctuations increase by a factor of seven near the wall in the interval $-0.02 \le x/L \le 0.05$), with the largest peak in the figure observed at the furthest downstream location shown. 
Clear changes in the anisotropy are also evident near the wall. While during the strong FPG and at the bump peak the streamwise fluctuations were dominating significantly over the other two directions, resulting in long and stretched streamwise-oriented structures, the normal Reynolds stresses become more balanced and isotropic during the partial retransition, although the streamwise direction still dominates. These trends in the span- and time-averaged profiles are consistent with the vorticity contours in figure~\ref{fig:vortAPG}.

\begin{figure}
\centering
  \includegraphics[width = 0.55\textwidth]{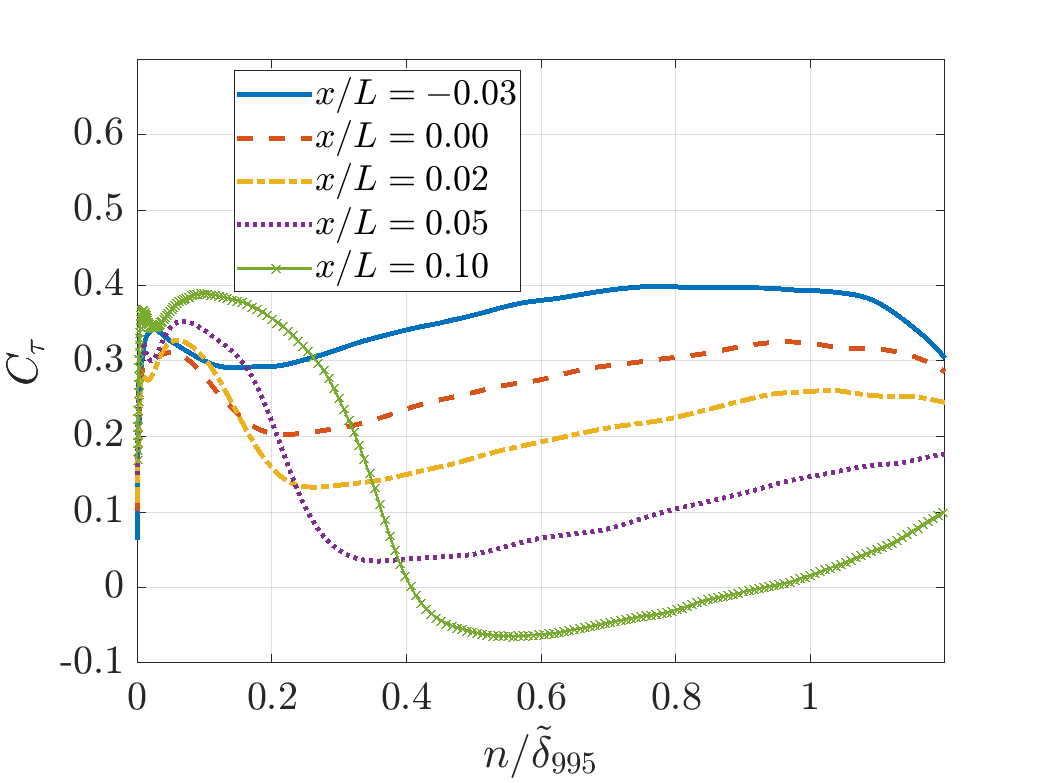}
  \caption{Correlation coefficient $C_\tau$ in the APG region of the Gaussian bump.}
\label{fig:CtauAPG}
\end{figure}

% Correlation
Figure~\ref{fig:CtauAPG} shows the correlation coefficient $C_\tau$ in the initial part of the strong APG. 
Relative to figure~\ref{fig:CtauFPG}, the profiles at the end of the FPG and the bump peak show a slight reduction in correlation near the wall, but a more substantial reduction in the outer layer due to the convex streamline curvature. As the boundary layer moves through the APG, the correlation gradually increases near the wall and spreads away from it. This is another indication of the second internal layer. In the outer layer, the $C_\tau$ keeps decreasing through the APG, reaching negative values by $x/L=0.1$ indicating that the streamwise and wall-normal fluctuations become positively correlated (note the negative sign \eqref{eq:Ctau}). These profiles are similar to those reported in \citet{Narasimha_Relaminarization_1979,SoMellor_Convex_73} and show the strength of the curvature effects, as well as the region wherein they are active in the boundary layer. 

\begin{figure}
\centering
  \includegraphics[width = 0.9\textwidth]{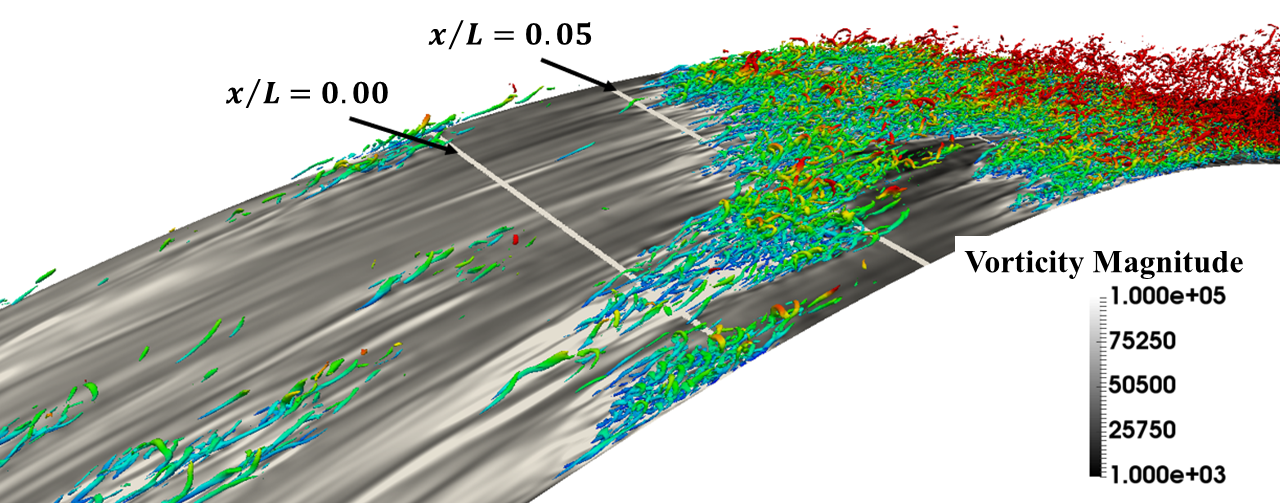}
  \caption{Isosurface of the instantaneous Q-criterion, $Q=6.0 \times 10^7$, and contours of vorticity magnitude over the peak and downstream side of the Gaussian bump.}
\label{fig:Qcrit}
\end{figure}

%% Revise summary
The peak of the bump and the APG region exhibit a partial retransition process that develops a second internal layer with significantly increased turbulent intensity and wall shear. This newly energized flow is more resilient to the strong APG, and thus only exhibits incipient flow separation on the downstream side of the bump. 
Pressure gradient effects are focused in the near-wall region and are the main drivers of the skin friction coefficient, whereas the outer layer is fairly independent and shows the continued action of convex streamline curvature.
A more three-dimensional view of the internal layer and the turbulent spots is shown in figure~\ref{fig:Qcrit} with an isosurface of the Q-criterion. The weakened near-wall turbulence and intermittent quiet regions at the bump peak are also visible with the contours of vorticity magnitude at the wall.
It must be stated once again that the formation of this internal layer is directly dependent on the upstream FPG effects, and thus the two regions of the flow are highly connected. 
A turbulence model that does not predict the weakening of the near-wall turbulence due to the FPG and instead models the flow as being fully turbulent everywhere will also not predict the partial retransition at the peak of the bump and consequently compute the wrong APG response and separation. This is the case of the SA model used for the preliminary simulations of the bump in figure~\ref{fig:CpCf}.
Alternatively, a model that does predict the FPG weakening effects but not the correct response at the bump peak will also not accurately compute the details of the separation. Adequate modeling of both of these physical phenomena is required for correct prediction of the effects of strong pressure gradients alternating in rapid succession as shown here, which is a common feature of aerodynamic flows.
%lift and drag forces for this flow.

% Production and budgets
\subsection{Reynolds Stress Production Rates}
\label{sec:ResProd}
% Discuss TKE and shear production and budgets

Additional interesting and useful quantities for the analysis of the boundary layer over the Gaussian bump are the production rates for the turbulent kinetic energy and the Reynolds shear stress.
These quantities are particularly useful to evaluate and improve RANS turbulence models.
%, some of which solve modeled equations for the TKE and the dissipation rate.

\begin{figure}
\centering
  \includegraphics[width = 0.97\textwidth]{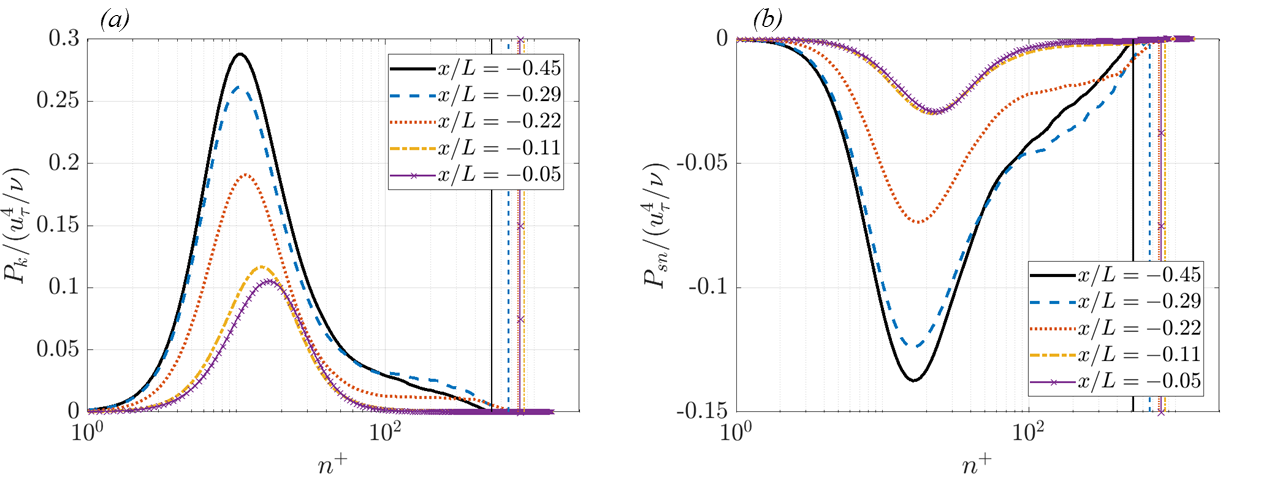}
  \caption{Production rate of the turbulent kinetic energy (a) and Reynolds shear stress (b) in the FPG region of the Gaussian bump non-dimensionalized by wall units. The vertical lines mark the boundary layer thickness $\tilde{\delta}_{995}$ for each streamwise location.}
\label{fig:prodFPGPlus}
\end{figure}

Figure~\ref{fig:prodFPGPlus} presents the production rates of the TKE and Reynolds shear stress over the upstream side of the bump. These quantities are defined as
\begin{equation}
P_k=-\overline{u'_iu'_j}\frac{\partial \overline{u}_i}{\partial x_j}
\qquad
\qquad
P_{sn}=-\left(  \overline{u'_su'_k}\frac{\partial \overline{u}_n}{\partial x_k}  +   \overline{u'_nu'_k}\frac{\partial \overline{u}_s}{\partial x_k}   \right),
\label{eq:prod}
\end{equation}
respectively, where Einstein notation is used to represent the double contraction of two tensors and the indices $i$, $j$, and $k$ take the three dimensions of the curvilinear coordinate system $(s,n,z)$. 
Note the non-dimensionalization by wall units, which allows a comparison of the production rates relative to the increase in wall shear due to the acceleration. Moreover, the distance to the wall is expressed in wall units to emphasize the near-wall region where these quantities are the largest.

At the start of the FPG, the profiles appear similar to the ones in the upstream mild APG, although production of both quantities is notably increased in the outer later due to the upstream APG and concave streamline curvature.
Immediately downstream of this location, a significant reduction in the magnitude of the inner layer peaks is observed. Given the destabilizing concave curvature of the streamlines upstream of $x/L=-0.14$, these changes are attributed to the strong pressure gradient. The peaks are also seen to move further from the wall in $n^+$ units, indicating once again the thickening of the viscous sublayer. 
The TKE production peak settles just above a non-dimensional value of 0.1 (approximately a factor of 2.5 smaller relative to the start of the FPG) and undergoes only a small change in the region $-0.11 \le x/L \le -0.05$ between the peak favorable pressure gradient and the skin friction peak. 
The Reynolds shear stress production follows similar trends, but exhibits a larger reduction of the inner peak relative to the start of the FPG (approximately a factor of five smaller).

Further from the wall in the outer layer ($n^+ \gtrsim 100$), both production rates in figure~\ref{fig:prodFPGPlus} decay rapidly and become negligible by $x/L=-0.11$. Once again, since convex curvature is present only downstream of $x/L=-0.14$, this reduction is mainly attributed to the strong FPG, although curvature effects are contributing past this location. The strong acceleration, therefore, effectively turns off production in the outer layer, suggesting once again how this region behaves as a free-shear layer with only a small dependence on the wall.
Figure~\ref{fig:prodFPGPlus} therefore highlights the effects of a relaminarization process through a weakening of the near-wall turbulent production relative to the rising wall shear and acceleration. The profiles, however, do not become negligible in magnitude, giving further evidence of the fact that the relaminarization process does not complete and the flow over the bump is still partially turbulent.

\section{Conclusions}
\label{sec:Conc}

Direct numerical simulation was performed of the turbulent boundary layer over a Gaussian shaped bump. The smooth surface causes a series of alternating pressure gradients and mean streamline curvature effects which combine to form a multitude of complex flow physics with significant deviation from standard turbulence behavior. %Incipient separation is also observed on the downstream side. 
%At the inflow to the DNS, the momentum thickness Reynolds number is approximately $1,000$ and the boundary layer thickness is $1/8$ of the bump height. 
The domain of interest for this study is the portion of the boundary layer from the inflow to the point of incipient separation, with particular emphasis on the strong favorable pressure gradient. 

Due to a strong acceleration, the boundary layer exhibits the stages of a relaminarization process. 
The standard logarithmic law breaks down soon after the start of the FPG where the pressure gradient and acceleration parameters are far below their critical values. % and profiles of the streamwise velocity deviate above it. This behavior is observed for small values of the pressure gradient and acceleration parameters, $\Delta_p=-0.013$ and $K=1.4 \times 10^{-6}$ respectively, relative to their critical values identified in previous studies. 
The non-dimensional shear stress parameter, however, 
%$\Delta_\tau$ and its critical value of $-0.013$ were 
was found to be predictive of the start of the deviation above the logarithmic law. 
Instantaneous contours of vorticity 
%at multiple heights above the wall 
indicate that the boundary layer is still fully turbulent where the velocity deviates above the logarithmic law, although the turbulent character and intensity are altered by the FPG.
%This result demonstrates that these pressure gradient effects alone are not a sign of relaminarization, but rather they precede the onset of this process.

Under continued acceleration, the near-wall region gradually ceases to be fully turbulent and intermittent spots of quiet flow with low vorticity grow in size. This is considered to be the onset of relaminarization, the process by which fully turbulent flow progresses towards a quasi-laminar state. 
This is in spite of the pressure and acceleration parameters never reaching the critical values of $\Delta_p=-0.025$ and $K=3 \times 10^{-6}$, respectively, suggested in previous studies to mark the onset of this process.
Relaminarization, however, does not complete and thus a quasi-laminar state is not achieved.
%Note that relaminarization continues until slightly upstream of the bump peak where the favorable pressure gradient relaxes, however the process does not complete and thus a quasi-laminar state is not achieved. 
In this region of the FPG, the Reynolds shear stress is reduced developing a bi-modal shape and the skin friction coefficient drops. When normalized by the local wall stress, the significant reduction of the turbulent kinetic energy, the Reynolds stresses, and their production rates clearly describes the weakening of the turbulence relative to the acceleration of the flow.
%Under this process, the streamwise velocity deviates more above the logarithmic law as the viscous layer grows thicker, the Reynolds shear stress is reduced and takes a bi-modal shape, and the skin friction coefficient drops. The near-wall peak of the streamwise component of the Reynolds stresses and the turbulent kinetic energy increase in absolute terms, but decrease significantly when normalized by the local wall stress (and even more so does the Reynolds shear stress with this normalization) clearly describing the weakening of the turbulence relative to the acceleration of the flow. Similarly, the production rates of the turbulent kinetic energy and Reynolds shear stress normalized by wall units are significantly reduced in the inner layer, and shut off altogether in the outer layer. 
%Furthermore, it is interesting to remark that the flow enters a relaminarization process in spite of $\Delta_p$ and $K$ only coming close, but never reaching, the critical values of $-0.025$ and $3 \times 10^{-6}$, respectively, suggested in previous studies to mark the onset of this process.
%By contrast to the near-wall region, the outer layer stays fully turbulent, although the fluctuations are significantly suppressed by the acceleration and convex curvature effects.

At the peak of the bump, as the pressure gradient changes to adverse, a partial retransition process is observed near the wall. The weakened flow from the upstream acceleration experiences a sudden enhancement in turbulent intensity, producing spots of large vorticity and a surge in turbulent kinetic energy and Reynolds shear stress. The skin friction coefficient is also increased forming a local maximum. The newly energized boundary layer is more resilient to the strong deceleration on the downstream side, only resulting in incipient separation.

Two internal layers are also formed in the region of interest, causing the inner and outer regions of the boundary layer to be largely independent of each other during the strong favorable and adverse pressure gradients. The flow near the wall is dominated by the pressure gradients and is responsible for the skin friction coefficient, while further from the wall the turbulence behaves similarly to a free-shear layer subject to pressure gradients and streamline curvature. Curvature effects are found to be negligible in the near-wall region.

%The Gaussian bump studied provides a novel numerical example of a complex flow undergoing relaminarization by a strong favorable pressure gradient followed by retransition. 
% Can I offer a series of events for strong FPG and relaminarization (maybe this goes at the end of the section so I can include things like the departure from the log law)?

RANS computations with the Spalart-Allmaras model of the same Gaussian bump flow resulted in a significant overprediction of the skin friction coefficient over most of the favorable pressure gradient and over the bump peak, as well as a much larger separation bubble due to the inability of the linear eddy viscosity closure to predict the correct pressure gradient effects. 
Similar results are observed with the SST $k$-$\omega$ model in the context of wall-modeled LES \citep{BalinAviation2020}. 
Future work is therefore focused on utilizing the DNS data for the improvement of RANS predictions of boundary layers under strong favorable pressure gradients. 
Moreover, wall-resolved large eddy simulations of the Gaussian bump at the same Reynolds number are in progress with the aim of validating a novel approach against the current DNS. This exercise will be followed by more LES of the same geometry at a twice larger the Reynolds number where the relaminarization is expected to be eliminated.

%%%%%%%%%%%%%%% Acknowledgements  %%%%%%%%%%%%%%%%%%%%%%%%
This work was supported by the National Science Foundation, Chemical, Bioengineering, Environmental and Transport Systems grant CBET-1710670 and by the National Aeronautics and Space Administration, Transformational Tools and Technologies grant 80NSSC18M0147, both to the University of Colorado Boulder. 
Computational resources were utilized at the NASA High-End Computing (HEC) Program through the NASA Advanced Supercomputing (NAS) Division at Ames Research Center and at the Argonne Leadership Computing Facility (ALCF), which is a DOE Office of Science User Facility supported under Contract DE-AC02-06CH11357. 
Finally, the authors thank Drs. P.R. Spalart, J.A. Evans, and M.K. Strelets for the helpful insight and communications regarding the problem setup, the analysis of the flow, and the synthetic turbulence generation method.
\newline \textbf{Declaration of Interests:} The authors report no conflict of interest.

%\appendix
%\section{}\label{appA}

\bibliographystyle{jfm}
\bibliography{biblio}

\begin{thebibliography}{51}
\expandafter\ifx\csname natexlab\endcsname\relax\def\natexlab#1{#1}\fi
\def\au#1{#1} \def\ed#1{#1} \def\yr#1{#1}\def\at#1{#1}\def\jt#1{\textit{#1}}
  \def\bt#1{#1}\def\bvol#1{\textbf{#1}} \def\vol#1{#1} \def\pg#1{#1}
  \def\publ#1{#1}\def\arxiv#1{#1}\def\org#1{#1}\def\st#1{\textit{#1}}

\bibitem[Abe(2017)]{Abe_DNS_2017}
{\sc \au{Abe, H.}} \yr{2017}  \at{Reynolds-number dependence of wall-pressure
  fluctuations in a pressure-induced turbulent separation bubble}.  \jt{Journal
  of Fluid Mechanics}  \bvol{833},  \pg{563--598}.

\bibitem[Abe {\em et~al.\/}(2012)Abe, Mizobuchi, Matsuo \&
  Spalart]{Abe_DNS_2012}
{\sc \au{Abe, H.}, \au{Mizobuchi, Y.}, \au{Matsuo, Y.} \& \au{Spalart, P.~R.}}
  \yr{2012}  \at{{DNS} and modeling of a turbulent boundary layer with
  separation and reattachment over a range of {Reynolds} numbers}.  \jt{Center
  for Turbulence Research Annual Research Briefs 2012}  \pg{pp. 311--322}.

\bibitem[Balin {\em et~al.\/}(2020)Balin, Jansen \& Spalart]{BalinAviation2020}
{\sc \au{Balin, R.}, \au{Jansen, K.~E.} \& \au{Spalart, P.~R.}} \yr{2020}
  \bt{Wall-modeled {LES} of flow over a {G}aussian bump with strong pressure
  gradients and separation}. AIAA 2020-3012.

\bibitem[Baskaran {\em et~al.\/}(1987)Baskaran, Smits \&
  Joubert]{Baskaran_part1}
{\sc \au{Baskaran, V.}, \au{Smits, A.~J.} \& \au{Joubert, P.~N.}} \yr{1987}
  \at{A turbulent flow over a curved hill. {Part} 1. {Growth} of an internal
  boundary layer}.  \jt{Journal of Fluid Mechanics}  \bvol{182},  \pg{47--83}.

\bibitem[Baskaran {\em et~al.\/}(1991)Baskaran, Smits \&
  Joubert]{Baskaran_part2}
{\sc \au{Baskaran, V.}, \au{Smits, A.~J.} \& \au{Joubert, P.~N.}} \yr{1991}
  \at{A turbulent flow over a curved hill. {Part} 2. {Effects} of streamline
  curvature and streamwise pressure gradient}.  \jt{Journal of Fluid Mechanics}
   \bvol{232},  \pg{377--402}.

\bibitem[Blackwelder \& Kovasznay(1972)]{Blackwelder72}
{\sc \au{Blackwelder, R.~F.} \& \au{Kovasznay, L. S.~G.}} \yr{1972}
  \at{Large-scale motion of a turbulent boundary layer during
  relaminarization}.  \jt{Journal of Fluid Mechanics}  \bvol{53},  \pg{61--83}.

\bibitem[Bradshaw(1969)]{Bradshaw1969}
{\sc \au{Bradshaw, P.}} \yr{1969}  \at{A note on reverse transition}.
  \jt{Journal of Fluid Mechanics}  \bvol{35},  \pg{387--390}.

\bibitem[Cal \& Castillo(2008)]{Cal08}
{\sc \au{Cal, R.~B.} \& \au{Castillo, L.}} \yr{2008}  \at{Similarity analysis
  of favorable pressure gradient turbulent boundary layers with eventual
  quasilaminarization}.  \jt{Physics of Fluids}  \bvol{20},  \pg{105106}.

\bibitem[Cavar \& Meyer(2011)]{Cavar2011}
{\sc \au{Cavar, D.} \& \au{Meyer, K.~E.}} \yr{2011}  \at{Investigation of
  turbulent boundary layer flow over {2D} bump using highly resolved large eddy
  simulation}.  \jt{Journal of Fluids Engineering}  \bvol{133},  \pg{111204}.

\bibitem[Coleman {\em et~al.\/}(2018)Coleman, Rumsey \&
  Spalart]{Coleman_DNS_2018}
{\sc \au{Coleman, G.~N.}, \au{Rumsey, C.~L.} \& \au{Spalart, P.~R.}} \yr{2018}
  \at{Numerical study of turbulent separation bubbles with varying pressure
  gradient and {Reynolds} number}.  \jt{Journal of Fluid Mechanics}
  \bvol{847},  \pg{28--70}.

\bibitem[Coles(1962)]{Coles1962}
{\sc \au{Coles, D.~E.}} \yr{1962}  \bt{The turbulent boundary layer in a
  compressible fluid}. {\em Tech. Rep.\/} R-403-PR.  \org{The Rand
  Corporation}.

\bibitem[Finnicum \& Hanratty(1988)]{Finnicum1988}
{\sc \au{Finnicum, D.~S.} \& \au{Hanratty, T.~J.}} \yr{1988}  \at{Effect of
  favorable pressure gradients on turbulent boundary layers}.  \jt{AIChE
  Journal}  \bvol{34},  \pg{529--540}.

\bibitem[Greenblatt {\em et~al.\/}(2006)Greenblatt, Paschal, Yao, Harris,
  Schaeffler \& Washburn]{Greenblatt_NASAhump}
{\sc \au{Greenblatt, D.}, \au{Paschal, K.~B.}, \au{Yao, C-S.}, \au{Harris, J.},
  \au{Schaeffler, N.~W.} \& \au{Washburn, A.~E.}} \yr{2006}  \at{Experimental
  investigation of separation control {Part} 1: {Baseline} and steady suction}.
   \jt{AIAA Journal}  \bvol{44},  \pg{2820--2830}.

\bibitem[Jansen {\em et~al.\/}(2000)Jansen, Whiting \&
  Hulbert]{Jansen_GenAlpha_2000}
{\sc \au{Jansen, Kenneth~E.}, \au{Whiting, Christian~H.} \& \au{Hulbert,
  Gregory~M.}} \yr{2000}  \at{{Generalized-$\alpha$ method for integrating the
  filtered Navier-Stokes equations with a stabilized finite element method}}.
  \jt{Computer Methods in Applied Mechanics and Engineering} .

\bibitem[Jimenez {\em et~al.\/}(2010)Jimenez, Hoyas, Simens \&
  Mizuno]{Jimenez_ZPGDNS}
{\sc \au{Jimenez, J.}, \au{Hoyas, S.}, \au{Simens, M.~P.} \& \au{Mizuno, Y.}}
  \yr{2010}  \at{Turbulent boundary layers and channels at moderate {Reynolds}
  numbers}.  \jt{Journal of Fluid Mechanics}  \bvol{657},  \pg{335--360}.

\bibitem[Kitsios {\em et~al.\/}(2016)Kitsios, Atkinson, Sillero, Borrell,
  Gungor, Jimenez \& Soria]{Kitsios_DNS_2016}
{\sc \au{Kitsios, V.}, \au{Atkinson, C.}, \au{Sillero, J.~A.}, \au{Borrell,
  G.}, \au{Gungor, A.~G.}, \au{Jimenez, J.} \& \au{Soria, J.}} \yr{2016}
  \at{Direct numerical simulation of a self-similar adverse pressure gradient
  turbulent boundary layer}.  \jt{International Journal of Heat and Fluid Flow}
   \bvol{61},  \pg{129--136}.

\bibitem[Manhart \& Friedrich(2002)]{Manhart_DNS_2002}
{\sc \au{Manhart, M.} \& \au{Friedrich, R.}} \yr{2002}  \at{{DNS} of a
  turbulent boundary layer with separation}.  \jt{International Journal of Heat
  and Fluid Flow}  \bvol{23},  \pg{572--581}.

\bibitem[Matai \& Durbin(2019)]{Matai_BumpLES_2019}
{\sc \au{Matai, R.} \& \au{Durbin, P.}} \yr{2019}  \at{Large-eddy simulation of
  turbulent flow over a parametric set of bumps}.  \jt{Journal of Fluid
  Mechanics}  \bvol{866},  \pg{503--525}.

\bibitem[Na \& Moin(1998)]{Na_DNS_1998}
{\sc \au{Na, Y.} \& \au{Moin, P.}} \yr{1998}  \at{Direct numerical simulation
  of a separated turbulent boundary layer}.  \jt{Journal of Fluid Mechanics}
  \bvol{374},  \pg{379--405}.

\bibitem[Narasimha \& Sreenivasan(1973)]{Narasimha1973}
{\sc \au{Narasimha, R.} \& \au{Sreenivasan, K.~R.}} \yr{1973}
  \at{Relaminarization in highly accelerated turbulent boundary layers}.
  \jt{Journal of Fluid Mechanics}  \bvol{61},  \pg{417--447}.

\bibitem[Narasimha \& Sreenivasan(1979)]{Narasimha_Relaminarization_1979}
{\sc \au{Narasimha, R.} \& \au{Sreenivasan, K.~R.}} \yr{1979}
  \at{Relaminarization of fluid flows}.  \jt{Advances in Applied Mechanics}
  \bvol{19},  \pg{222--303}.

\bibitem[Narayanan \& Ramjee(1969)]{BadriNarayanan1969}
{\sc \au{Narayanan, M. A.~Badri} \& \au{Ramjee, V.}} \yr{1969}  \at{On the
  ctiteria for reverse transition in a two-dimensional boundary layer flow}.
  \jt{Journal of Fluid Mechanics}  \bvol{35},  \pg{225--241}.

\bibitem[Patel(1965)]{Patel_FPG}
{\sc \au{Patel, V.~C.}} \yr{1965}  \at{Calibration of the {Preston} tube and
  limitations on its use in pressure gradients}.  \jt{Journal of Fluid
  Mechanics}  \bvol{23},  \pg{185--208}.

\bibitem[Patel \& Head(1968)]{Patel_ReverseTrans_1968}
{\sc \au{Patel, V.~C.} \& \au{Head, M.~R.}} \yr{1968}  \at{Reversion of
  turbulent to laminar flow}.  \jt{Journal of Fluid Mechanics}  \bvol{34},
  \pg{371--392}.

\bibitem[Schlatter {\em et~al.\/}(2009)Schlatter, \"Orl\"u, Li, Brethouwer,
  Fransson, Johansson, Alfredsson \& Henningson]{Schlatter2009}
{\sc \au{Schlatter, P.}, \au{\"Orl\"u, R.}, \au{Li, Q.}, \au{Brethouwer, G.},
  \au{Fransson, J. H.~M.}, \au{Johansson, A.~V.}, \au{Alfredsson, P.~H.} \&
  \au{Henningson, D.~S.}} \yr{2009}  \at{Turbulent boundary layers up to
  ${R}e_\theta=2500$ studied through simulation and experiment}.  \jt{Physics
  of Fluids}  \bvol{21},  \pg{051702}.

\bibitem[Schwarz \& Plesniak(1996)]{Schwarz1996}
{\sc \au{Schwarz, A.C.} \& \au{Plesniak, M.~W.}} \yr{1996}  \at{Convex
  turbulent boundary layers with zero and favorable pressure gradients}.
  \jt{Journal of Fluids Engineering}  \bvol{118},  \pg{787--794}.

\bibitem[Shin \& Song(2015)]{Shin2015}
{\sc \au{Shin, J.~H.} \& \au{Song, S.~J.}} \yr{2015}  \at{Pressure gradient
  effects on smooth- and rough-surface turbulent boundary layers -- part 1:
  Favorable pressure gradient}.  \jt{Journal of Fluids Engineering}
  \bvol{137},  \pg{011203}.

\bibitem[Shur {\em et~al.\/}(2014)Shur, Spalart, Strelets \& Travin]{Shur_STG}
{\sc \au{Shur, M.~L.}, \au{Spalart, P.~R.}, \au{Strelets, M.~Kh.} \&
  \au{Travin, A.~K.}} \yr{2014}  \at{Synthetic turbulence generators for
  {RANS}-{LES} interfaces in zonal simulations of aerodynamic and aeroacoustic
  problems}.  \jt{Flow, Turbulence and Combustion}  \bvol{93},  \pg{63--92}.

\bibitem[Shur {\em et~al.\/}(2000)Shur, Strelets, Travin \& Spalart]{Shur_SARC}
{\sc \au{Shur, M.~L.}, \au{Strelets, M.~K.}, \au{Travin, A.~K.} \& \au{Spalart,
  P.~R.}} \yr{2000}  \at{Turbulence modeling in rotating and curved channels:
  {Assessing} the {Spalart}-{Shur} correction}.  \jt{AIAA Journal}  \bvol{38},
  \pg{784--792}.

\bibitem[Skote \& Henningson(2002)]{Skote_DNS_2002}
{\sc \au{Skote, M.} \& \au{Henningson, D.~S.}} \yr{2002}  \at{Direct numerical
  simulation of a separated turbulent boundary layer}.  \jt{Journal of Fluid
  Mechanics}  \bvol{471},  \pg{107--136}.

\bibitem[Slotnik(2019)]{Slotnik_SpeedBump_2019}
{\sc \au{Slotnik, J.~P.}} \yr{2019}  \at{Integrated {CFD} validation
  experiments for prediction of turbulent separated flows for subsonic
  transport aircraft}.  \jt{STO-MP-AVT-307} .

\bibitem[Smits {\em et~al.\/}(1983)Smits, Matheson \& Joubert]{Smits1983}
{\sc \au{Smits, A.~J.}, \au{Matheson, N.} \& \au{Joubert, P.~N.}} \yr{1983}
  \at{Low-{R}eynolds-number turbulent boundary layers in zero and favorable
  pressure gradients}.  \jt{Journal of Ship Research}  \bvol{27},  \pg{147}.

\bibitem[So \& Mellor(1973)]{SoMellor_Convex_73}
{\sc \au{So, R. M.~C.} \& \au{Mellor, G.~L.}} \yr{1973}  \at{Experiment on
  convex curvature effects in turbulent boundary layers}.  \jt{Journal of Fluid
  Mechanics}  \bvol{60},  \pg{43--62}.

\bibitem[So \& Mellor(1975)]{SoMellor_Concave_75}
{\sc \au{So, R. M.~C.} \& \au{Mellor, G.~L.}} \yr{1975}  \at{Experiment on
  turbulent boundary layers on a concave wall}.  \jt{Aeronautical Quarterly}
  \bvol{26},  \pg{25--40}.

\bibitem[Spalart(1986)]{Spa86}
{\sc \au{Spalart, P.R.}} \yr{1986}  \at{Numerical study of sink-flow boundary
  layers}.  \jt{Journal of Fluid Mechanics}  \bvol{172},  \pg{307--328}.

\bibitem[Spalart \& Allmaras(1994)]{spalart1994one}
{\sc \au{Spalart, P.~R.} \& \au{Allmaras, S.~R.}} \yr{1994}  \at{A one-equation
  turbulence model for aerodynamic flows}.  \jt{Recherche Aerospatiale}
  \bvol{1},  \pg{5--21}.

\bibitem[Spalart {\em et~al.\/}(2017)Spalart, Belyaev, Garbaruk, Shur, Strelets
  \& Travin]{Spalart_BJWMLES}
{\sc \au{Spalart, P.~R.}, \au{Belyaev, K.~V.}, \au{Garbaruk, A.~V.}, \au{Shur,
  M.~L.}, \au{Strelets, M.~Kh.} \& \au{Travin, A.~K.}} \yr{2017}
  \at{Large-eddy and direct numerical simulations of the {Bachalo}-{Johnson}
  flow with shock-induced separation}.  \jt{Flow, Turbulence and Combustion}
  \bvol{99},  \pg{865--885}.

\bibitem[Spalart \& Coleman(1997)]{Spalart_DNS_1997}
{\sc \au{Spalart, P.~R.} \& \au{Coleman, G.~N.}} \yr{1997}  \at{Numerical study
  of a separation bubble with heat transfer}.  \jt{European Journal of
  Mechanics -- B/Fluids}  \bvol{16},  \pg{169--189}.

\bibitem[Spalart \& Shur(1997)]{Spalart_SARC}
{\sc \au{Spalart, P.~R.} \& \au{Shur, M.~L.}} \yr{1997}  \at{On the
  sensitization of turbulence models to rotation and curvature}.  \jt{Aerospace
  Science and Technology}  \bvol{5},  \pg{297--302}.

\bibitem[Spalart \& Watmuff(1993)]{Spalart_DNS_1993}
{\sc \au{Spalart, P.~R.} \& \au{Watmuff, J.~H.}} \yr{1993}  \at{Experimental
  and numerical study of a turbulent boundary layer with pressure gradients}.
  \jt{Journal of Fluid Mechanics}  \bvol{249},  \pg{337--371}.

\bibitem[Sreenivasan(1982)]{Sreenivasan82}
{\sc \au{Sreenivasan, K.~R.}} \yr{1982}  \at{Laminarescent, relaminarizing and
  retransitional flows}.  \jt{Acta Mechanica}  \bvol{44},  \pg{1--48}.

\bibitem[Trofimova {\em et~al.\/}(2009)Trofimova, Tejada-Martinez \&
  Jansen]{Trofimova_DNS_2009}
{\sc \au{Trofimova, A.~V.}, \au{Tejada-Martinez, A.~E.} \& \au{Jansen, K.E.}}
  \yr{2009}  \at{Direct numerical simulation of turbulent channel flows using a
  stabilized finite element method}.  \jt{Computers \& Fluids}  \bvol{38},
  \pg{924--938}.

\bibitem[Tsuji \& Morikawa(1976)]{Tsuji_hills}
{\sc \au{Tsuji, Y.} \& \au{Morikawa, Y.}} \yr{1976}  \at{Turbulent boundary
  layer with pressure gradient alternating in sign}.  \jt{The Aeronautical
  quarterly}  \bvol{27},  \pg{15--28}.

\bibitem[Uzun \& Malik(2018)]{Uzun_NASAHumpLES_2018}
{\sc \au{Uzun, A.} \& \au{Malik, M.~R.}} \yr{2018}  \at{Large-eddy simulation
  of flow over a wall-mounted hump with separation and reattachment}.  \jt{AIAA
  Journal}  \bvol{56},  \pg{715--730}.

\bibitem[Uzun \& Malik(2020)]{Uzun_SpeedBumpDNS_2020}
{\sc \au{Uzun, A.} \& \au{Malik, M.~R.}} \yr{2020} Simulation of a turbulent
  flow subjected to favorable and adverse pressure gradients.  \bt{In {\em AIAA
  AVIATION Forum\/}}. Virtual Event.

\bibitem[Warnack \& Fernholz(1998)]{Warnack1998}
{\sc \au{Warnack, D.} \& \au{Fernholz, H.~H.}} \yr{1998}  \at{The effects of a
  favorable pressure gradient and of the {R}eynolds number on an incompressible
  axisymmetric turbulent boundary layer. {P}art 2. {T}he boundary layer with
  relaminarization}.  \jt{Journal of Fluid Mechanics}  \bvol{359},
  \pg{357--381}.

\bibitem[Webster {\em et~al.\/}(1996)Webster, DeGraaff \& Eaton]{Webster_bump}
{\sc \au{Webster, D.~R.}, \au{DeGraaff, D.~B.} \& \au{Eaton, J.~K.}} \yr{1996}
  \at{Turbulence characteristics of a boundary layer over a two-dimensional
  bump}.  \jt{Journal of Fluid Mechanics}  \bvol{320},  \pg{53--69}.

\bibitem[Whiting \& Jansen(1999)]{Whiting1999}
{\sc \au{Whiting, Christian~H.} \& \au{Jansen, Kenneth~E.}} \yr{1999}
  \at{Stabilized finite element methods for fluid dynamics using a hierarchical
  basis}. PhD thesis, Rensselaer Polytechnic Institute.

\bibitem[Williams {\em et~al.\/}(2020)Williams, Samuell, Sarwas, Robbins \&
  Ferrante]{Williams_ExpSpeedBump}
{\sc \au{Williams, Owen}, \au{Samuell, Madeline}, \au{Sarwas, E.~Sage},
  \au{Robbins, Matthew} \& \au{Ferrante, Antonino}} \yr{2020} Experimental
  study of a {CFD} validation test case for turbulent separated flows.  \bt{In
  {\em AIAA Scitech 2020 Forum\/}}. Orlando.

\bibitem[Wright {\em et~al.\/}(2020)Wright, Balin, Patterson, Evans \&
  Jansen]{WrightSTGDNS_arxiv}
{\sc \au{Wright, J.}, \au{Balin, R.}, \au{Patterson, J.~W.}, \au{Evans, J.~A.}
  \& \au{Jansen, K.~E.}} \yr{2020}  \at{Direct numerical simulation of a
  turbulent boundary layer on a flat plate using synthetic turbulence
  generation}.  \jt{arXiv:2010.3407543v1 [physics.flu-dyn]} .

\bibitem[Wu \& Squires(1998)]{Wu_LESbump}
{\sc \au{Wu, X.} \& \au{Squires, K.~D.}} \yr{1998}  \at{Numerical investigation
  of the turbulent boundary layer over a bump}.  \jt{Journal of Fluid
  Mechanics}  \bvol{362},  \pg{229--271}.

\end{thebibliography}

\end{document}